\documentclass[aps,prd,twoside,onecolumn,superscriptaddress,floatfix,nofootinbib,showpacs]{revtex4}
\usepackage{amsmath}
\usepackage{bm}
\usepackage{xcolor}

\newcommand\beq{\begin{equation}}
\newcommand\eeq{\end{equation}}
\newcommand\beqn{\begin{eqnarray}}
\newcommand\eeqn{\end{eqnarray}}

\newcommand\bl{{\bm{\ell}}}

\newcommand\nn{\nonumber}

\def\x{{\bf x}}

\def\k{{\bf k}}

\def\q{{\bf q}}
\def\d{{\bf d}}

\def\bl{{\bf L}}
\def\L{{ \bm \ell}}

\def\kc{{\kappa_{\rm CMB}}}
\def\kg{{\kappa_{\rm gal}}}

\newcommand{\ba}{\begin{eqnarray}}
\newcommand{\ea}{\end{eqnarray}}
\newcommand{\be}{\begin{equation}}
\newcommand{\ee}{\end{equation}}

\def\tTheta{\tilde{\Theta}}

\newcommand\lsim{\mathrel{\rlap{\lower4pt\hbox{\hskip1pt$\sim$}}
        \raise1pt\hbox{$<$}}}
\newcommand\gsim{\mathrel{\rlap{\lower4pt\hbox{\hskip1pt$\sim$}}
        \raise1pt\hbox{$>$}}}

\newcommand{\jcap}{{J.~Cosm.~Astrop.~Phys.}}
\newcommand{\araa}{{Annu.~Rev.~Astron.~Astrophys.}}
\newcommand{\aap}{{Astron.~Astrophys.}}
\newcommand{\apjl}{{Astrophys.~J.~Lett.}}
\newcommand{\apjs}{{Astrophys.~J.~Supp.}}

\newcommand{\mnras}{{Mon.~Not.~R.~Astron.~Soc.}}
\usepackage{graphicx}
\usepackage[large]{subfigure}
\usepackage{amssymb, amsmath}
\usepackage[amssymb]{SIunits}
\usepackage{epstopdf}
\usepackage{aas_macros}
\usepackage{natbib}

\begin{document}

\title{Bias to CMB Lensing Reconstruction from Temperature Anisotropies due to Large-Scale Galaxy Motions}

 \author{Simone~Ferraro}
 \affiliation{Berkeley Center for Cosmological Physics and
Department of Astronomy, University of California, Berkeley, CA, USA 94720}
  \affiliation{Miller Institute for Basic Research in Science, University of California, Berkeley, CA, 94720 USA}
  \author{J.~Colin~Hill}
 \affiliation{Dept.~of Astronomy, Pupin Hall, Columbia University, New York, NY USA 10027}
\begin{abstract}
Gravitational lensing of the cosmic microwave background (CMB) is expected to be amongst the most powerful cosmological tools for ongoing and upcoming CMB experiments.  In this work, we investigate a bias to CMB lensing reconstruction from temperature anisotropies due to the kinematic Sunyaev-Zel'dovich (kSZ) effect, that is, the Doppler shift of CMB photons induced by Compton-scattering off moving electrons.  The kSZ signal yields biases due to both its own intrinsic non-Gaussianity and its non-zero cross-correlation with the CMB lensing field (and other fields that trace the large-scale structure). This kSZ-induced bias affects both the CMB lensing auto-power spectrum and its cross-correlation with low-redshift tracers.  Furthermore, it cannot be removed by multifrequency foreground separation techniques because the kSZ effect preserves the blackbody spectrum of the CMB.  While statistically negligible for current datasets, we show that it will be important for upcoming surveys, and failure to account for it can lead to large biases in constraints on neutrino masses or the properties of dark energy.  For a Stage 4 CMB experiment,
the bias can be as large as $\approx$ 15\% or 12\% in cross-correlation with LSST galaxy lensing convergence or galaxy overdensity maps, respectively, when the maximum temperature multipole used in the reconstruction is $\ell_{\rm max} = 4000$, and about half of that when $\ell_{\rm max} = 3000$. Similarly, we find that the CMB lensing auto-power spectrum can be biased by up to several percent.
These biases are many times larger than the expected statistical errors. We validate our analytical predictions with cosmological simulations and present the first complete estimate of secondary-induced CMB lensing biases. The predicted bias is sensitive to the small-scale gas distribution, which is affected by pressure and feedback mechanisms, thus making removal via ``bias-hardened'' estimators challenging. Reducing $\ell_{\rm max}$ can significantly mitigate the bias at the cost of a decrease in the overall lensing reconstruction signal-to-noise. A bias $\lesssim 1$\% on large scales requires $\ell_{\rm max} \lesssim 2000$, which leads to a reduction in signal-to-noise by a factor of $\approx 3-5$ for a Stage 4 CMB experiment.  Polarization-only reconstruction may be the most robust mitigation strategy.
\end{abstract}
\maketitle

\section{Introduction}
Matter inhomogeneities between our location and the surface of last scattering deflect cosmic microwave background (CMB) photons, introducing new correlations in the observed CMB anisotropies. These correlations allow the projected gravitational potential sourced by the late-time matter distribution to be extracted from high-resolution maps of the microwave sky, a procedure known as CMB lensing reconstruction. CMB lensing probes the density field over a wide range of redshifts ($0.5 \lesssim z \lesssim 6$) and is dominated by contributions from linear modes for angular scales up to multipole $L \approx 1000$.  It is therefore an excellent probe of dark energy, modified gravity, and the sum of the neutrino masses \cite{2006PhR...429....1L, 2015APh....63...66A}.

The CMB lensing power spectrum will be measured with signal-to-noise ($S/N$) $> 100$ by ongoing and upcoming experiments, including the Advanced Atacama Cosmology Telescope (AdvACT)~\cite{2016JLTP..184..772H}, the South Pole Telescope-3G (SPT-3G)~\cite{2014SPIE.9153E..1PB}, the Simons Observatory\footnote{\url{http://www.simonsobservatory.org/}}, and CMB Stage-4\footnote{\url{http://www.cmb-s4.org/}} (CMB-S4)~\cite{2016arXiv161002743A}.  At this level of precision, sub-percent control is required on possible biases in CMB lensing reconstruction.  Such biases can result from both instrumental or astrophysical effects; here, we will focus on the latter.  In particular, the observed CMB temperature fluctuations are a sum of the lensed primary fluctuations (which alone would give rise to an unbiased lensing reconstruction, modulo estimator-related complexities~\cite{Kesden2003,Hanson2011,2016PhRvD..94d3519B}) and several secondary anisotropies due to the interaction (either gravitational or electromagnetic) of CMB photons with late-time structures.  These secondary anisotropies include the thermal and kinematic Sunyaev-Zel'dovich (SZ) effects \cite{Zeldovich-Sunyaev1969,Sunyaev-Zeldovich1970,Sunyaev-Zeldovich1972, Sunyaev-Zeldovich1980}, the integrated Sachs-Wolfe (ISW) effect~\cite{1967ApJ...147...73S}, and the non-linear generalization of the ISW effect known as the Rees-Sciama effect~\cite{1968Natur.217..511R}.  In addition, the microwave sky includes signals from thermal dust emission (both Galactic and extragalactic) and radio emission, which must be carefully treated in CMB analyses. While most of the secondary and astrophysical signals can be separated from the lensed primary CMB using multifrequency component separation methods, such procedures cannot isolate the kinematic SZ (kSZ) and ISW effects, since they preserve the blackbody spectrum of the CMB.\footnote{Relativistic corrections to the kSZ effect~(e.g.,~\cite{Nozawa2006}) generate a non-blackbody frequency dependence, but are negligible for the purposes of our analysis.} The linear ISW effect is only relevant on large angular scales ($\gtrsim 1$ degree), making it straightforward to filter out in the lensing reconstruction process if needed, while the Rees-Sciama effect is expected to be roughly two orders of magnitude smaller than the kSZ signal on the relevant scales \cite{2009PhRvD..80f3528S, 2002PhRvD..65h3518C}. We therefore focus on the kSZ-induced bias, being the largest among the effects that cannot be mitigated by multifrequency component separation.\footnote{There are also CMB lensing reconstruction biases due to the non-Gaussianity of the late-time matter field~\cite{2016PhRvD..94d3519B}, which are by definition blackbody in frequency-dependence, but these are distinct from the secondary anisotropy-induced biases. }  Imperfect removal of non-blackbody foregrounds can also lead to significant biases in CMB lensing reconstruction, as has been explored in detail for the thermal SZ and dusty galaxy (cosmic infrared background [CIB]) signals~\cite{2014ApJ...786...13V, 2014JCAP...03..024O}, as well as for the polarized dust emission from our Galaxy~\cite{2012JCAP...12..017F}.\footnote{Note that Refs.~\cite{2014ApJ...786...13V, 2014JCAP...03..024O} primarily focused on single-frequency measurements, but at much higher noise levels than considered in this paper, thus yielding a kSZ-induced bias much smaller than the statistical uncertainties (as we show explicitly for the Planck SMICA map in Sec.~\ref{sec:results}) and much smaller than the unmasked tSZ- or CIB-induced biases.}

The kSZ effect is a Doppler shift due to the Compton-scattering of CMB photons off of free electrons moving with a non-zero line-of-sight (LOS) velocity~\cite{Sunyaev-Zeldovich1972,Sunyaev-Zeldovich1980,Ostriker-Vishniac1986}.  The corresponding shift in the observed CMB temperature is proportional to the total number of electrons and their LOS velocity, i.e., the LOS electron momentum.  Being a Doppler shift, the kSZ effect preserves the blackbody spectrum of the CMB, leading to only a small change in the blackbody temperature (to lowest order). The kSZ signal can be used to measure the ionized gas abundance and distribution in galaxies and clusters, thus providing important information about the extent and nature of astrophysical feedback processes (e.g., energy injection from active galactic nucleus feedback).  Taking advantage of this sensitivity to the gas distribution, recent detections of the kSZ signal have made progress towards resolving the long-standing ``missing baryons'' problem at low redshift~\cite{Handetal2012,Planck2016kSZ,Schaan:2015uaa,2016arXiv160301608H,2016MNRAS.461.3172S}.  In addition, through its dependence on the large-scale velocity field, the kSZ effect can also be used as a cosmological probe to measure the growth of structure~\cite{Mueller2015a,Mueller2015b,Alonso2016}. In this paper, however, we will focus on the bias it imprints on CMB lensing reconstruction.

Building on previous work~\cite{Doreetal2004, DeDeoetal2005}, it was shown in \cite{2016arXiv160301608H, 2016arXiv160502722F} that the kSZ signal can be efficiently extracted by cross-correlating the \textit{square} of an appropriately filtered CMB temperature map with a sample of large-scale structure tracers.  In contrast to other kSZ estimators, this method does not require spectroscopic redshift information for the tracer sample, relying only on the projected tracer distribution, thus allowing kSZ measurements with densely-sampled photometric surveys. In \cite{2016arXiv160301608H} the first kSZ detection using this method was achieved using the Planck, WMAP, and Wide-field Infrared Survey Explorer (WISE) datasets. These analyses also noted that because this kSZ estimator is quadratic in CMB temperature, it is significantly contaminated by the CMB lensing signal.  The CMB lensing contribution was detected at high significance and had to be marginalized over in order to obtain a reliable kSZ measurement.  Turning the problem around, one would thus expect a contribution from the kSZ signal to the quadratic lensing reconstruction estimator.  Quantifying this bias is the focus of this paper.  To our knowledge, this effect has only been investigated in detail in~\cite{2004NewA....9..687A}, who found that it could lead to biases of order unity on the reconstructed CMB lensing power spectrum for a low-noise ($2 \, \mu$K-arcmin), high-resolution (0.8 arcmin) experiment.  The effect was also discussed briefly in~\cite{2012ApJ...756..142V}, who found a sub-percent bias to the CMB lensing auto-power spectrum for higher-noise maps ($18 \, \mu$K-arcmin) of similar resolution ($\approx 1$ arcmin).  Given the dramatic increase in our knowledge of the microwave sky in recent years, as well as the expected precision of upcoming CMB experiments, it is timely to revisit this issue.

The CMB lensing power spectrum is a sensitive probe of the amplitude of fluctuations at relatively low redshift, probing the integrated growth of structure between recombination and $z = 0$, with a broad peak around $z \approx 2$.  Thus, it is a probe of the constituents of the Universe. For example, massive neutrinos produce a few-percent suppression of the CMB lensing power spectrum compared to a cosmology with massless neutrinos~\cite{2015APh....63...66A, 2012MNRAS.425.1170H} (with the amount of suppression being proportional to the neutrino mass sum). Therefore, even percent-level biases in the lensing power spectrum can yield large biases on cosmological parameters of interest.  Moreover, cross-correlations of CMB lensing maps with galaxy overdensity or galaxy weak lensing maps also directly probe the late-time growth of structure, providing a powerful test of gravity and dark energy models \cite{1999ApJ...522L..21H}, as well as calibration of systematics \cite{2016arXiv160701761S}.

While the kSZ signal affects only CMB temperature and not polarization fluctuations (to lowest order), the statistical power of CMB lensing reconstruction will be dominated by temperature in the next generation of CMB surveys, and will represent a statistically non-negligible contribution even for experiments for which the overall reconstruction is dominated by polarization, such as the proposed CMB-S4 survey.  Thus, although polarization-only reconstruction allows the kSZ-induced bias to be avoided, the consequence could be a significant decrease in the overall lensing $S/N$, depending on the experimental configuration. 

In our analysis we assume a flat $\Lambda$CDM fiducial cosmology with Planck 2015 parameters (column 3 of Table 4 of \cite{2016A&A...594A..13P}). We also assume massless neutrinos in the fiducial model, and compare the size of the kSZ-induced bias to the effect of minimal mass, normal hierarchy neutrinos in Section \ref{subsec:results:auto}.

The remainder of this paper is organized as follows:  In Section \ref{sec:kSZ} we review the kSZ effect, and in Section \ref{sec:lensrec} we review the process of lensing reconstruction from CMB temperature anisotropy measurements. The kSZ-induced bias to the cross-correlation between CMB lensing and low-redshift tracers is explored in Section \ref{sec:cross}, while the effect on the CMB lensing auto-power spectrum is investigated in Section \ref{sec:auto}. In Section \ref{sec:results} we show numerical estimates of the bias for upcoming surveys.  In Section \ref{sec:sims} we test the approximations made by comparison with cosmological simulations, which we also use to perform a complete calculation of the bias to the lensing auto-power spectrum (modulo reionization contributions). We consider mitigation strategies in Section~\ref{sec:mitigation} and conclude in Section \ref{sec:conclusions}.  Detailed derivations of the main results of the paper are found in Appendices \ref{app:cross} and \ref{app:auto}.

\section{The kinematic SZ effect}
\label{sec:kSZ}
The kSZ effect produces a CMB temperature change, $\Theta^{\rm kSZ}(\hat{\mathbf{n}}) = \Delta T^{\rm kSZ}(\hat{\mathbf{n}})/T_{\rm CMB}$, in a direction $\hat{\mathbf{n}}$ on the sky (in units with $c$ = 1):
\ba
\label{eq.kSZdef}
\Theta^{\rm kSZ}(\hat{\mathbf{n}}) & = & - \int d\eta \ g(\eta) \ \mathbf{p}_e \cdot \mathbf{\hat{n}} \\
& = & - \sigma_T \int \frac{d \eta}{1+z} e^{-\tau} n_e(\hat{\mathbf{n}},\eta) \ \mathbf{v}_e \cdot \mathbf{\hat{n}} \,,
\label{eq.kSZdef2}
\ea
where $\sigma_T$ is the Thomson scattering cross-section, $\eta(z)$ is the comoving distance to redshift $z$, $\tau$ is the optical depth to Thomson scattering, $g(\eta) = (d \tau / d\eta) e^{-\tau}$ is the visibility function, $n_e$ is the physical free electron number density, $\mathbf{v}_e$ is the peculiar velocity of the electrons, and we have defined the electron momentum $\mathbf{p}_e = (1+\delta_e) \mathbf{v}_e$.  The sign has been chosen such that electrons with positive LOS velocity produce a negative kSZ signal.

Significant kSZ anisotropies are produced in cosmological epochs during which there are large fluctuations in electron density. Such fluctuations are present at late times in galaxies and clusters due to the non-linear growth of structure, and also earlier during the epoch of reionization, where fluctuations in the electron density field are due to fluctuations in the ionization fraction \cite{2013ApJ...776...83B, 2013ApJ...769...93P, 2016ApJ...824..118A}. While the latter are also expected to be correlated with the matter density field and hence with CMB lensing, they are located at $z \gtrsim 7$.  Due to the declining geometric kernel for CMB lensing at these high redshifts, their influence is likely much smaller in our analysis than the kSZ fluctuations at low redshift.  For this reason, we focus on the kSZ signal arising from late-time structures and defer a study of the effects of reionization to future work.  Nevertheless, our results should be taken as a lower limit on the kSZ-related biases in CMB lensing reconstruction (particularly for the auto-power spectrum), since reionization-generated kSZ fluctuations will also contribute at some level.

We also note that to lowest order in velocity, the kSZ effect produces only temperature and not polarization fluctuations.  Therefore, we will only consider lensing reconstruction from CMB temperature maps in this analysis.

\section{Lensing reconstruction from temperature fluctuations}
\label{sec:lensrec}
Gravitational lensing of the primary CMB introduces statistical correlations between different Fourier modes, which would otherwise be uncorrelated (under the hypothesis that the primordial fluctuations are a statistically isotropic Gaussian random field). These correlations allows the lensing field to be reconstructed from observed CMB maps, as we will describe.

In the absence of foregrounds, the observed, lensed fluctuations $\tTheta_p(\x)$ are related to the unlensed, primordial fluctuations $\Theta_p$ by a remapping under the \textit{displacement} field $\d(\x)$ \cite{2006PhR...429....1L}:
\be
\tTheta_p(\x) = \Theta_p(\x + \d)
\ee
To lowest order, it can be shown that the vector field $\d(\x)$ is irrotational, and therefore all of the information is contained in its divergence. For this reason, we will work with the CMB lensing convergence, conventionally defined as $\kc = -\frac12 \nabla \cdot \d$.  Physically, the CMB lensing convergence is a weighted projection of the matter density field back to the surface of last scattering (see Equations~\ref{eq.kappaCMBdef} and~\ref{eq.WkappaCMBdef} below).  It can then be shown that the minimum variance quadratic estimator for $\kc$ can be written as \cite{2006PhR...429....1L, 2002ApJ...574..566H}\footnote{For compactness, we use the notation $\int_{\L} \equiv \int \frac{d^2 \L}{(2\pi)^2}$ and $\int_{\eta} \equiv \int d\eta$.  Hats denote estimators for a given quantity. Upper-case $\mathbf{L}$ denotes lensing multipole, while lower-case $\L$ denotes temperature map multipole.  We assume the flat-sky approximation throughout.}:
\be
\hat{\kappa}_{\rm CMB}(\bl) = \frac{L^2 N(\bl)}{2} \int_{\L} \tTheta(\L) \tTheta(\bl - \L) f(\L, \bl) = \int_{\L} \tTheta(\L) \tTheta(\bl - \L) \Gamma(\L, \bl) \,,
\ee
where we have defined 
\be
\Gamma(\L, \bl) = \frac12 L^2 N(\bl) f(\L, \bl) 
\ee
and the mode-coupling kernel is
\be
f(\L, \bl) = \frac{ (\bl - \L) \cdot \bl C^{TT}_{|\bl - \L|} + \L \cdot \bl C^{TT}_{\ell} }{ 2C^{\rm tot}_{\ell}  C^{\rm tot}_{|\bl - \L|} }\ \ .
\ee
The reconstruction noise serves as the normalization in the estimator and represents the uncertainty in the reconstruction of $2 \kc(\bl)/L^2$ due to chance correlations between different modes in an unlensed, Gaussian realization:
\be
 N(\bl)^{-1} = \int_\L \frac{ \left[(\bl - \L) \cdot \bl C^{TT}_{|\bl - \L|} + \L \cdot \bl C^{TT}_{\ell} \right]^2 }{ 2C^{\rm tot}_{\ell}  C^{\rm tot}_{|\bl - \L|} } \ \ \ \ .
\ee
In all of the above, $\tTheta$ is the observed temperature fluctuation field, which is the sum of the lensed primordial fluctuations $\tTheta_p$ and the kSZ fluctuations\footnote{We will ignore the lensing of the kSZ fluctuations in this work.} $\Theta^{\rm kSZ}$, as well as detector noise with power spectrum $N^{\rm det}_\ell$, uncorrelated with all of the other components and given by
\be
N^{\rm det}_{\ell} =  \Delta^2_T e^{\theta^2_{\rm FWHM} \ell^2/(8 \ln2)} \,,
\ee
where $\Delta^2_T$ is the noise level of the experiment (usually quoted in $\mu$K-arcmin) and $\theta_{\rm FWHM}$ is the full-width at half-maximum (FWHM) of the beam in radians.

We will assume that all non-blackbody foregrounds have been removed by component separation and that the ISW fluctuations can be removed by filtering out scales $\gtrsim 1$ degree from the observed temperature map.  In practice, we use scales down to $\ell_{\rm min} = 30$, but increasing this cutoff to 100 or 200 would have negligible impact on our work, as there is effectively no accessible lensing or kSZ signal on these scales.  Throughout, $C^{\rm tot}_{\ell}$ denotes the \textit{total} power spectrum of the observed $\tTheta$, including the lensed primary CMB, kSZ, and detector noise.

\section{Bias to cross-correlation with large-scale structure tracers}
\label{sec:cross}
In this section, we investigate the kSZ-induced bias to the cross-correlation between low-redshift tracers (e.g., galaxies, quasars, or galaxy weak lensing convergence) and $\kc$.  We assume that $\kc$ is reconstructed from a temperature map containing kSZ and lensed primary fluctuations (as well as detector noise).  When considering galaxies or quasars as tracers, we define the projected tracer overdensity $\delta_g$ as
\be
\delta_g (\hat{\mathbf{n}}) =  \int_0^{\eta_{\rm max}} d \eta \ W^{g}(\eta) \ \delta_m(\eta \hat{\mathbf{n}}, \eta) \,,
\label{eq:delta_gproj}
\ee
where $\eta_{\rm max}$ is the maximum source distance, $\delta_m$ is the (three-dimensional) matter overdensity, and $W^g(\eta)$ is the projection kernel:
\be
W^{g} (\eta) = b_g p_s(\eta) \,.
\label{eq.Wg}
\ee
Here $p_s(\eta) \propto dn/d\eta$ is the distribution of the tracers in comoving distance (normalized to have unit integral) and $b_g$ is the linear tracer bias, which is allowed to be redshift-dependent.  

When considering galaxy lensing (i.e., cosmic shear) as a tracer, the convergence field $\kg$ is given by
\be
\kg (\hat{\mathbf{n}}) =  \int_0^{\eta_{\rm max}} d \eta\ W^{\kg}(\eta) \ \delta_m(\eta \hat{\mathbf{n}}, \eta) \,,
\label{eq.kappadef}
\ee
where $W^{\rm \kg}(\eta)$ is the lensing kernel:
\be
W^{\kg} (\eta) = \frac{3 \Omega_m H_0^2 \eta}{2 a} \int_\eta^{\eta_{\rm max}} d\eta_s \  p_s(\eta_s) \frac{\eta_s - \eta}{\eta_s} \,,
\label{eq.Wkappadef}
\ee
where $a$ is the scale factor and $p_s(\eta) \propto dn/d\eta$ is the distribution of sources in comoving distance (normalized to have unit integral).  For concreteness, we will use galaxies as tracers in the following, but all of the equations also hold for galaxy lensing with the replacement $W^g \rightarrow W^{\kg}$.

The lensed temperature fluctuations can be decomposed as the sum of a (lensed) primary component, the kSZ component, and noise: $\tTheta = \tTheta_p + \Theta^{\rm kSZ} + N^{\rm det}$.  Schematically, the cross-correlation of $\hat{\kappa}_{\rm CMB}$ with galaxies can be written as $\langle \delta_g \ \hat{\kappa}_{\rm CMB} \rangle \sim \langle \delta_g \ \tTheta \tTheta \rangle$, which can be expanded in $\tTheta_p$ and $\Theta^{\rm kSZ}$, yielding terms of the form  $\langle \delta_g \ \tTheta_p \tTheta_p \rangle$, $\langle \delta_g \ \tTheta_p \Theta^{\rm kSZ} \rangle$, and  $\langle \delta_g \ \Theta^{\rm kSZ} \Theta^{\rm kSZ} \rangle$. The first term gives the cross-correlation with the \textit{true} convergence field $\kappa_{\rm CMB}$, the second term vanishes on average, due to the equal probability of the kSZ signal being positive or negative, and the third term represents the bias to the cross-correlation arising from kSZ leakage into the CMB lensing reconstruction estimator.  Therefore, to lowest order, to calculate the kSZ-induced bias to CMB lensing cross-correlations, we can just replace $\tTheta \rightarrow \Theta^{\rm kSZ}$ in the $\hat{\kappa}_{\rm CMB}$ estimator.  

A computation outlined in Appendix \ref{app:cross} shows that the bias to the cross-correlation between $\kc$ and galaxy overdensity is given by:
\be
\left( \Delta  C_L^{\kc \times g} \right)_{\rm kSZ} = \int_{\eta} \frac{W^g(\eta) g^2(\eta)}{\eta^2} \mathcal{B}(\k = \bl / \eta; \eta)
\label{eq:cross}
\ee
with
\be
\mathcal{B}(\k; \eta) = \int_\q \Gamma(\k \eta + \q \eta, \k \eta) B_{\delta p_z p_z}(\k, \q, -\k - \q; \eta) \,.
\label{eq:trianglePS}
\ee
The hybrid bispectrum appearing in Equation~\ref{eq:trianglePS}, $B_{\delta p_z p_z}$, is the bispectrum of one power of the matter density field $\delta_m$ and two powers of the electron momentum projected along the line-of-sight, $p_z$. It has been shown that on scales smaller than the coherence length of the velocity field, the following is a good approximation \cite{Doreetal2004, DeDeoetal2005}:
\be
B_{\delta p_{z} p_{z} }\approx \frac{1}{3} v^2_{\rm rms} B_{\delta e e}^{\rm NL} \,,
\label{bispectrum}
\ee
where $v^2_{\rm rms}$ is the 3D velocity dispersion and $B_{\delta e e}^{\rm NL}$ is the non-linear bispectrum of matter and electron overdensity. As a first approximation, we can assume that the electrons trace the matter on the scales of interest and approximate $B_{\delta e e}^{\rm NL}$ with the non-linear matter bispectrum $B_m^{\rm NL}$. We will revisit this assumption in Section \ref{subsec:baryons}. Throughout, we use fitting functions from \cite{GilMarin:2011ik} for the non-linear matter bispectrum $B_m^{\rm NL}$, and the velocity dispersion $v^2_{\rm rms}$ is computed in linear theory, which has been shown to be an excellent approximation~\cite{Hahn:2014lca}.  We will compare the prediction of Equation~\ref{eq:cross} to cosmological simulations in Section \ref{sec:sims}.

At late times, a small fraction of the cosmological abundance of electrons lies in stars or neutral media and thus does not participate in the Thomson scattering that produces the kSZ signal. We define $f_{\rm free}$ as the fraction of free electrons, which is in general a function of redshift. The visibility function $g(\eta)$ in Equation~\ref{eq.kSZdef} is proportional to $f_{\rm free}$, so that $\left( \Delta C_L^{\kc \times g} \right)_{\rm kSZ} \propto f_{\rm free}^2$.  In the following, we will take $f_{\rm free} = 0.85$ as our fiducial value (except where stated otherwise), and note that $f_{\rm free}$ can be constrained with kSZ measurements~\cite{2016arXiv160701769S, 2016arXiv160502722F, Schaan:2015uaa, 2016MNRAS.461.3172S}.

\section{Bias to the CMB lensing power spectrum}
\label{sec:auto}
Similar to the calculation in the previous section, we also compute an analogous kSZ-induced bias to the reconstructed CMB lensing power spectrum, $\langle \hat{\kappa}_{\rm CMB} \hat{\kappa}_{\rm CMB} \rangle$. The relation between CMB lensing convergence and the underlying matter density field is given by setting the source for lensing to the surface of last scattering in Equation~\ref{eq.Wkappadef} (i.e., $p_s(\eta) = \delta_D(\eta - \eta_*)$, where $\eta_*$ is the comoving distance to recombination):
\be
\kc (\hat{\mathbf{n}}) =  \int_0^{\eta_*} d \eta\ W^{\kc}(\eta) \ \delta_m(\eta \hat{\mathbf{n}}, \eta) \,,
\label{eq.kappaCMBdef}
\ee
where $W^{\kc}(\eta)$ is the lensing kernel:
\be
W^{\kc} (\eta) = \frac{3 \Omega_m H_0^2 \eta}{2 a} \  \frac{\eta_* - \eta}{\eta_*} \,.
\label{eq.WkappaCMBdef}
\ee

The computation is greatly simplified by noting that $\kc$ is also a tracer of low-redshift structure (just like galaxies or galaxy lensing), and therefore, at the very least, there must be a bias to the CMB lensing power spectrum which is obtained by using Equation \ref{eq:cross} with the replacement $W^g \rightarrow W^\kc$ (and a combinatorial factor of 2, representing whether we consider the first or the second $\kc$ in $\langle \kc \ \kc \rangle$ as the tracer).  We calculate the kSZ-induced bias to the $\hat{\kappa}_{\rm CMB}$ auto-power spectrum in Appendix \ref{app:auto}, finding that the result can be approximated as 
\be
\left(\Delta C_L^{\kc} \right)_{\rm kSZ} \approx 2  \int_{\eta} \frac{W^\kc(\eta) g^2(\eta)}{\eta^2} \mathcal{B}(\k = \bl / \eta; \eta) + \text{(other terms)} \,,
\label{eq:auto}
\ee
where $\mathcal{B}(\k = \bl / \eta; \eta)$ is given by Equation~\ref{eq:trianglePS}.
The first term corresponds to the contribution discussed above (treating $\kc$ as a tracer of low-redshift structure). The ``other terms'' arise from different contractions of the fields, and include contributions from trispectra of the kSZ and ISW fields.  These terms are given in Appendix~\ref{app:auto}.  The contribution from the kSZ trispectrum was first investigated in \cite{2011PhRvL.107b1301D} and found to be negligible for the noise levels of the original ACT survey.  We include its contribution for Planck, CMB-S3, and CMB-S4 noise levels in the full simulation calculation presented in Sec~\ref{sec:fullauto}.   Note that for the tSZ and CIB-induced lensing biases, the trispectrum-induced bias on large scales was found to have the opposite sign to the term discussed above, and thus lead to partial cancellation in the overall bias \cite{2014ApJ...786...13V}.  Similarly, ``secondary contractions'' of the term in Eq.~\ref{eq:auto}, as described in Appendix \ref{app:auto}, may be of similar magnitude~\cite{2014JCAP...03..024O}.  We will compare the prediction of Equation~\ref{eq:auto} to cosmological simulations in Section \ref{sec:validation}, and will present the full result from simulations (including secondary contractions and the trispectrum) in Section \ref{sec:fullauto}.

\section{Numerical results for current and upcoming surveys}
\label{sec:results}
\subsection{Experimental configurations}
\label{subsec:configs}
We consider three idealized CMB experiments, summarized in Table \ref{tab:CMBconfig}:\footnote{Note that residual Poisson sources may increase the effective high-$\ell$ noise level over the white noise levels specified here, but the size of this effect depends sensitively on the source flux masking threshold and detailed experimental configuration (e.g., frequency coverage).} one with characteristics similar to the recent Planck SMICA component-separated map \cite{2016A&A...594A...9P}, one similar to the nominal specifications of ongoing Stage-3 experiments (which we will denote by CMB-S3) such as AdvACT \cite{2016JLTP..184..772H}, and finally a CMB-S4-like experiment~\cite{2016arXiv161002743A}.\footnote{The configuration for the proposed CMB-S4 experiment has not yet been set; therefore this case is for illustration purposes only.}  We consider lensing reconstruction from temperature anisotropies only and choose a multipole range from $\ell_{\rm min} = 30$ to $\ell_{\rm max} = 4000$ or $\ell_{\rm max} = 3000$ for the reconstruction in our fiducial analysis. We will further explore the effects of using a different multipole range for the lensing reconstruction in Section \ref{sec:mitigation}.

\begin{table}[h]
\begin{center}
  \begin{tabular}{| c | c | c |}
    \hline 
     CMB experiment & white noise level  & beam FWHM  \\ 
     \ &$\Delta_T$  [$\mu$K-arcmin] & $\theta_{\rm FWHM}$ [arcmin] \\ \hline \hline
    Planck SMICA & 45 & 5   \\ \hline
    CMB-S3 & 7 & 1.4   \\ \hline
    CMB-S4 & 1 & 3  \\ \hline
  \end{tabular}
  \caption{Experimental configurations considered in this work.}
  \label{tab:CMBconfig}
\end{center}
\end{table}

For low-redshift tracer samples, we consider galaxy density and galaxy lensing convergence maps extracted from Large Synoptic Survey Telescope (LSST) data.  We assume the following source distribution for the LSST ``gold'' sample with $i$-band magnitude $i < 25.3$ (\cite{2009arXiv0912.0201L}, Chapter 3):
\be
p_s(z) = \frac{1}{2z_0} \left( \frac{z}{z_0} \right)^2 e^{-z/z_0} \,,
\ee
where $z_0 = 0.0417 i - 0.744$.  We assume a linear galaxy bias of the form $b_g(z) = 1 + 0.84z$ (\cite{2009arXiv0912.0201L}, Chapter 13). The ``gold'' sample has a median redshift of $z_m \approx 0.8$ and galaxies extending out to $z \approx 3$.  We use this sample both as a galaxy number density sample and as a source sample for galaxy weak lensing convergence. For the forecasts involving LSST, we will assume that the shape noise is $\sigma_{\epsilon} = 0.26$ and the source number density $n = 26$ arcmin$^{-2}$.

\begin{figure}[ht]
\centerline{\includegraphics[width=10cm]{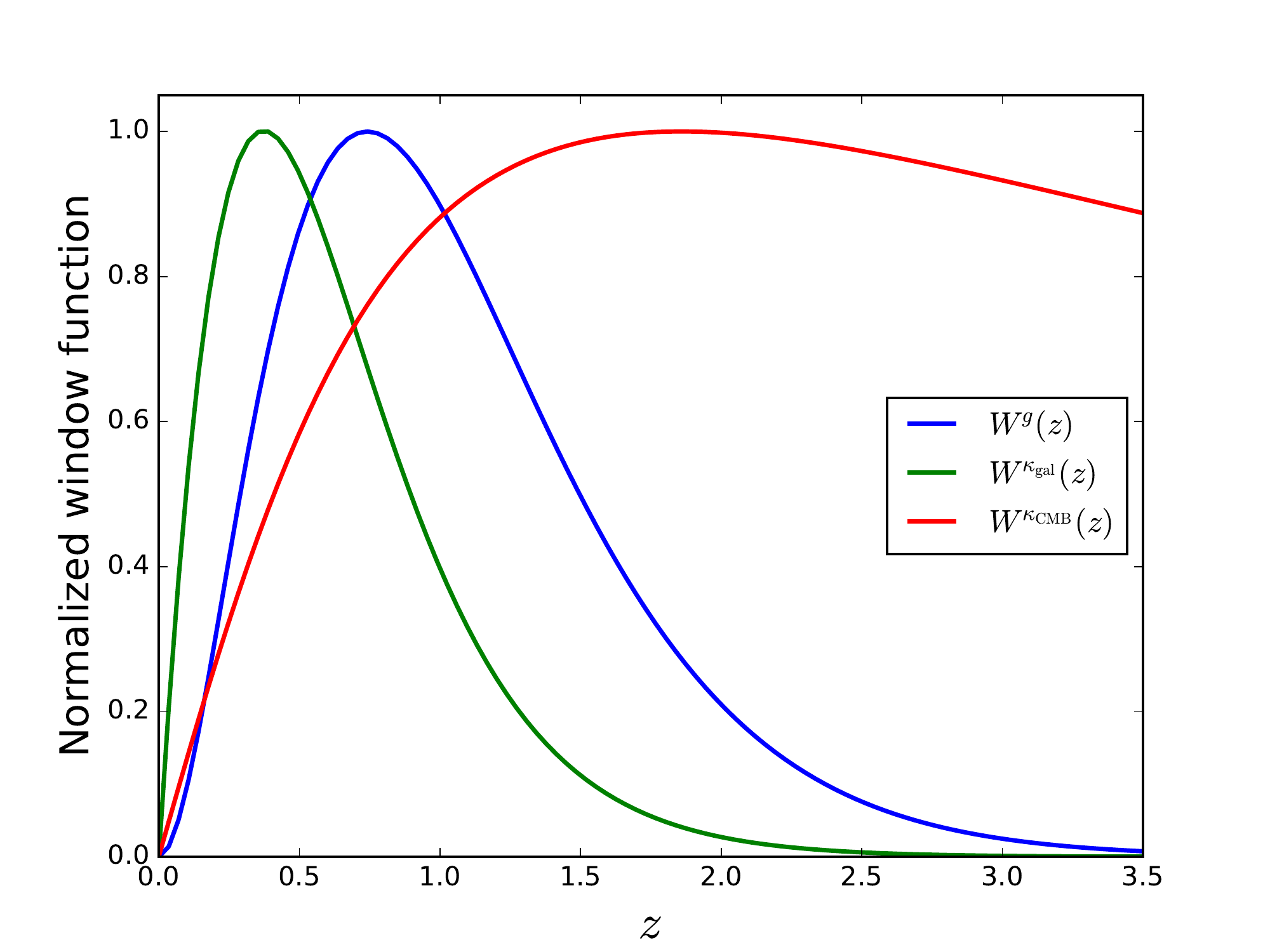}}
\caption{Window functions for LSST galaxy lensing (left peak), LSST galaxies (middle peak), and CMB lensing (right peak). They are related to the window functions defined in the text by $W(z) = W(\eta) d\eta/dz$.}
\label{fig:windows}
\end{figure}

The normalized window functions for LSST galaxies, LSST galaxy lensing, and CMB lensing are shown in Figure~\ref{fig:windows}.

\subsection{Baryonic physics}
\label{subsec:baryons}
In our fiducial model for the kSZ signal, we assume that the baryons trace the dark matter on the scales of interest.  This assumption is known to fail on small scales due to the effects of feedback and pressure support in galaxies and clusters. A full treatment of these baryonic processes requires high-resolution hydrodynamical simulations and is beyond the scope of this paper.  However, here we study the effect of pressure support, assuming that feedback acts on similar scales. Semianalytical models of gas dynamics predict that the gas overdensity $\delta_{\rm gas}$ is suppressed compared to the dark matter $\delta_{\rm cdm}$ below the \textit{filtering scale} $k_F$ \cite{2003ApJ...583..525G}:
\be
\delta_{e}(k,z) \approx \delta_{\rm gas}(k,z)  \approx \delta_{\rm cdm}(k,z) e^{-k^2 / k_F^2(z)} \,.
\label{eq:filtering}
\ee

The filtering scale $k_F$ is a time-integral of the Jeans scale $k_J(z) = a(z) \sqrt{4 \pi G \bar{\rho}_m} /c_s(z) $ that takes redshift evolution into account (here $c_s$ is the sound speed):
\be
\frac{1}{k_F^2(t)} = \frac{1}{D_{+}(t)} \int_0^t dt' \frac{\ddot{D}_+(t') + 2H(t')\dot{D}_+(t') } {k_J^2(t')} \ a^2(t') \int_{t'}^t \frac{dt''}{a^2(t'')} \,,
\label{eq:kJ}
\ee
where $D_+(t)$ is the linear growth factor.

In order to assess the impact of the baryon distribution on our results, we adopt an exponential suppression of the form in Equation \ref{eq:filtering}, and compare with the case in which baryons trace the dark matter.

\subsection{Results: cross-correlation with tracers}
\begin{figure}[ht]
\centering
\begin{tabular}{cc}
  \includegraphics[width=8.5cm]{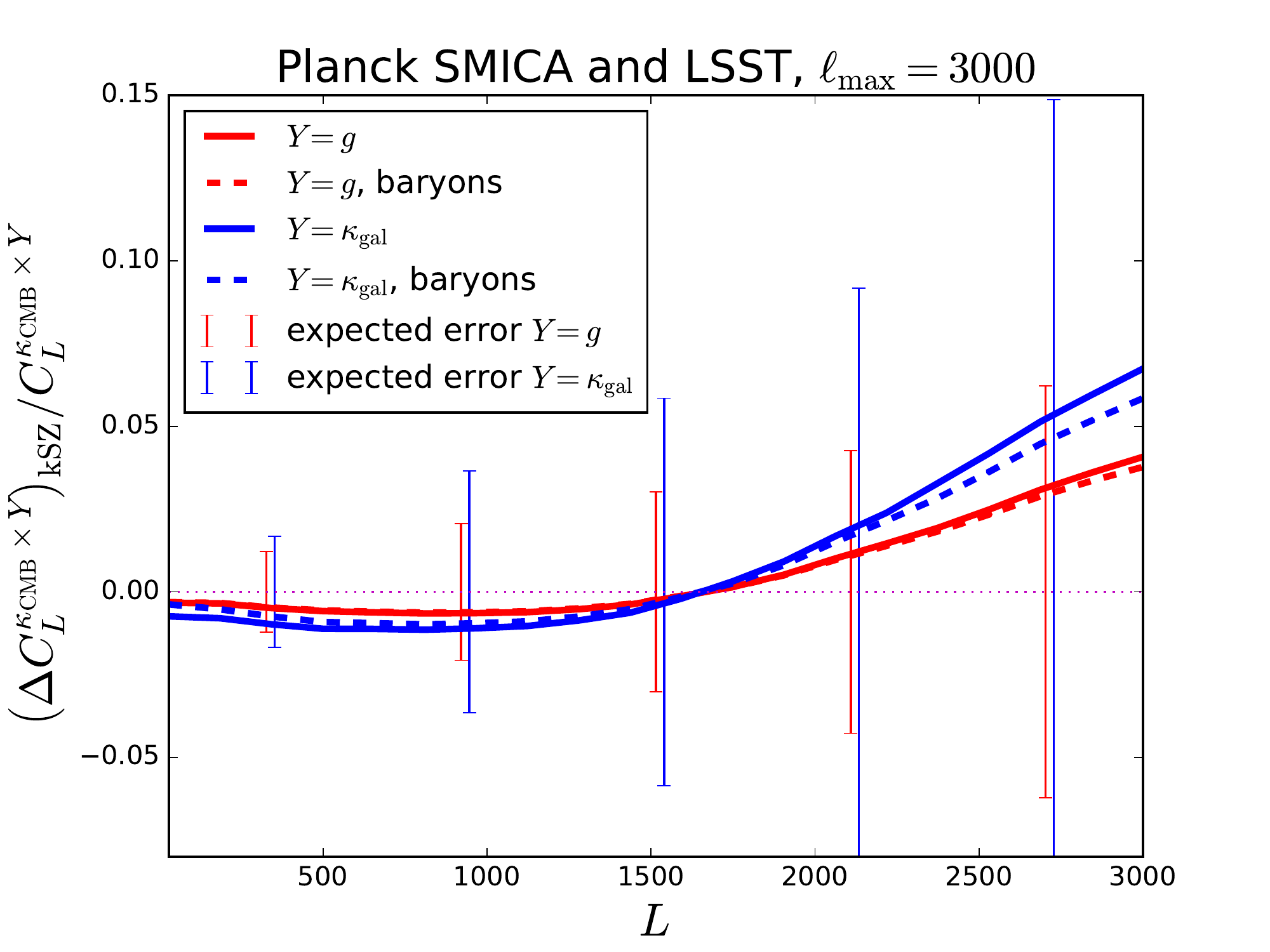} &   \includegraphics[width=8.5cm]{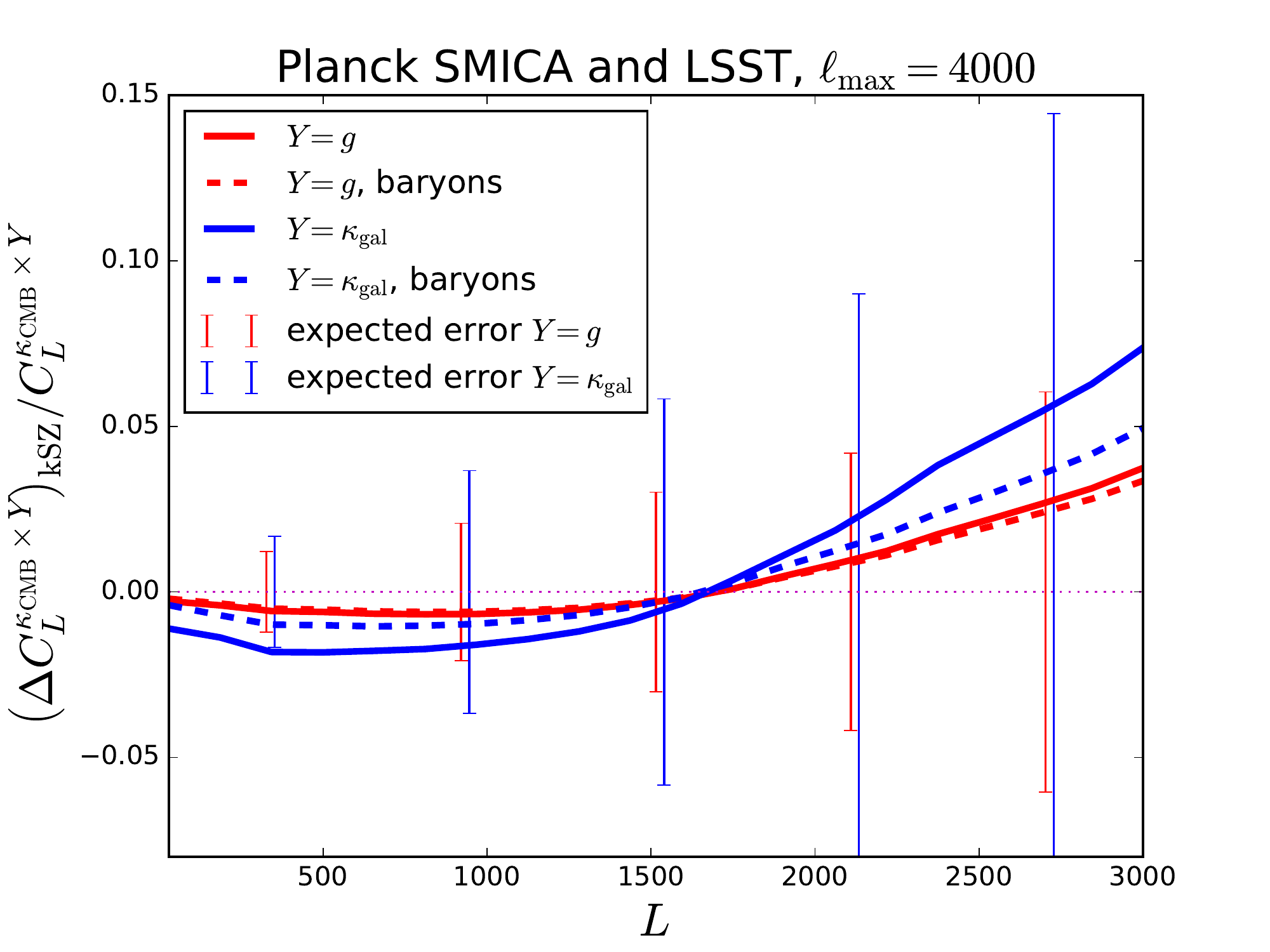} \\
  \includegraphics[width=8.5cm]{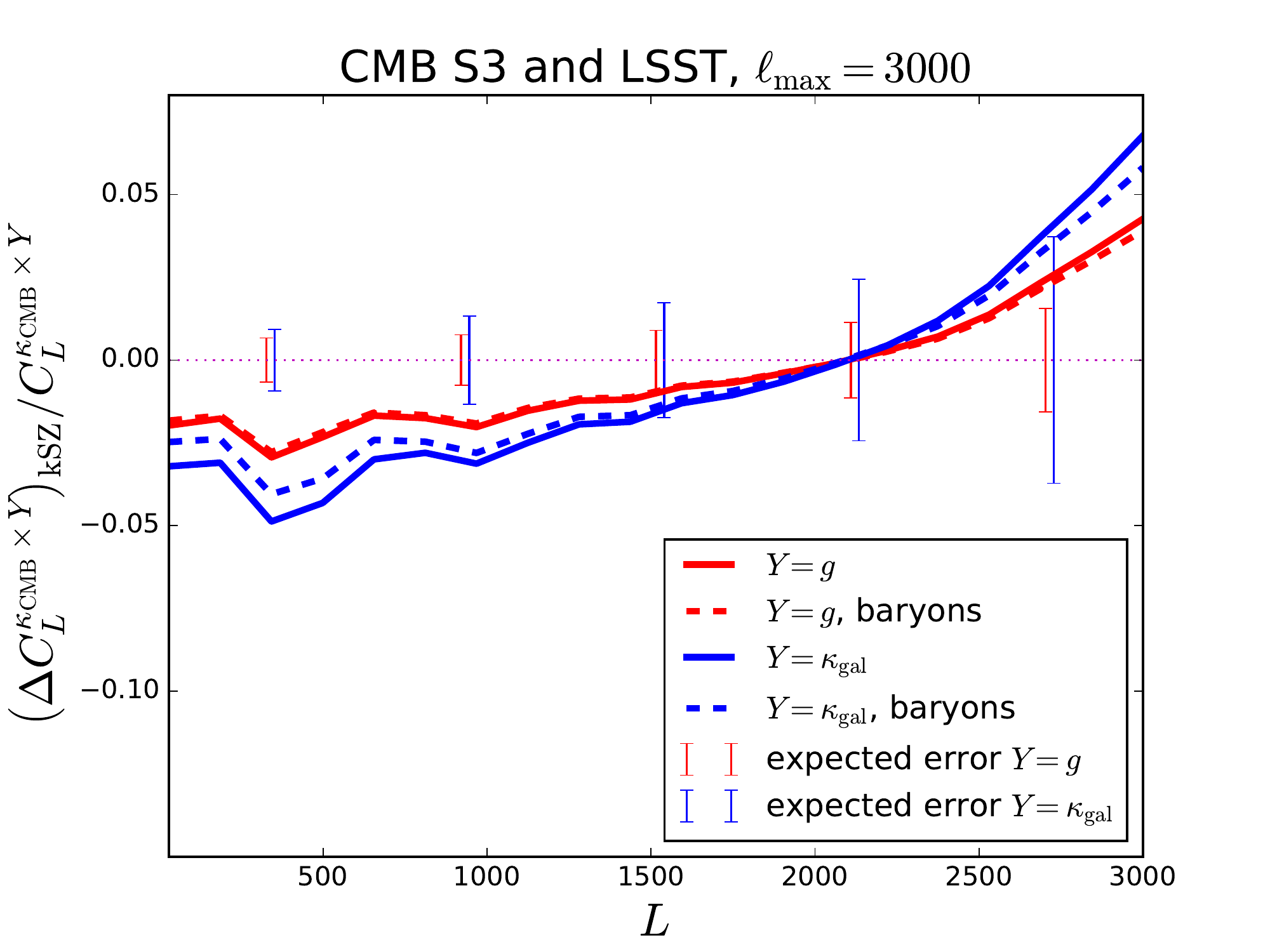} &   \includegraphics[width=8.5cm]{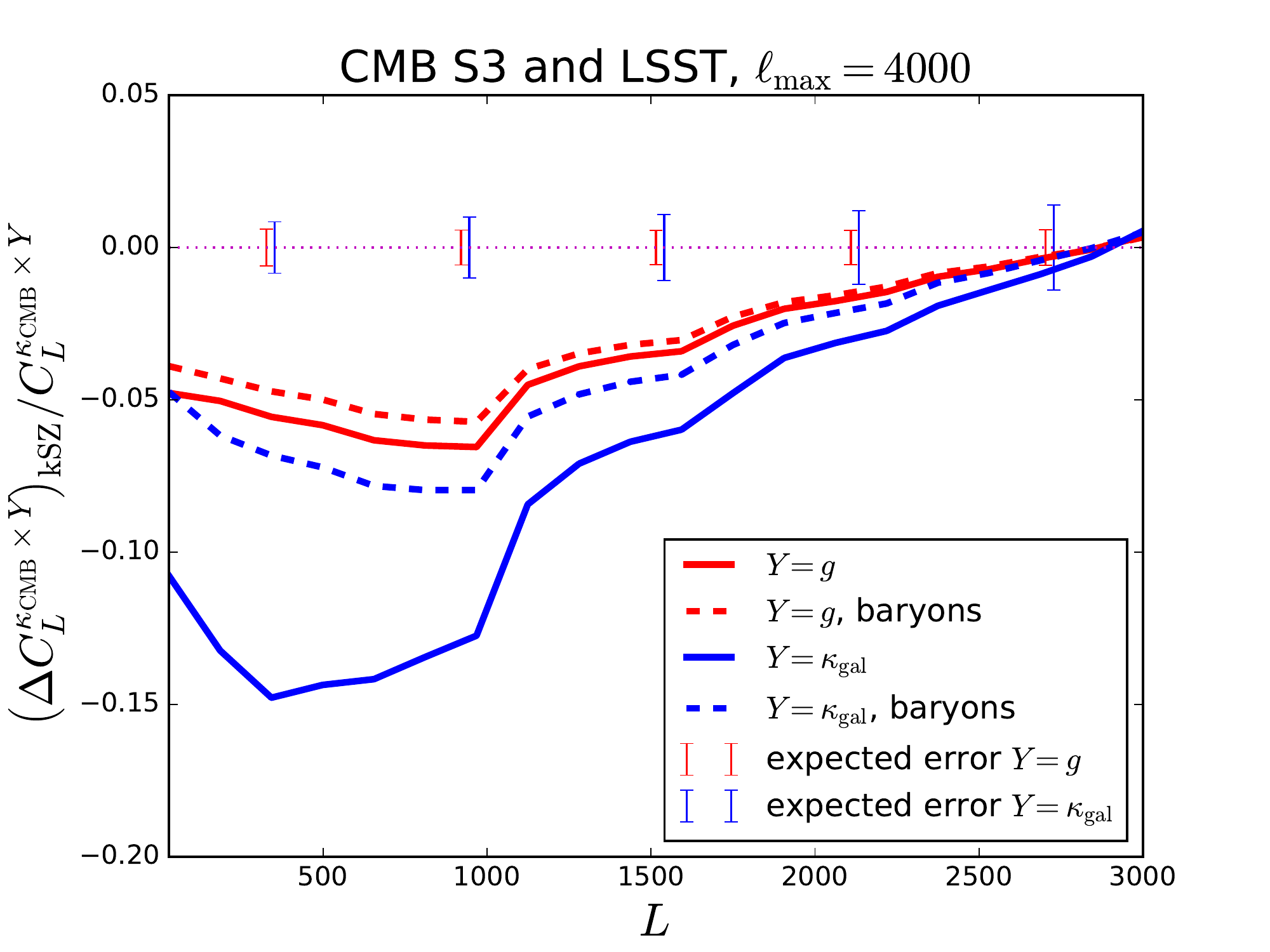} \\
  \includegraphics[width=8.5cm]{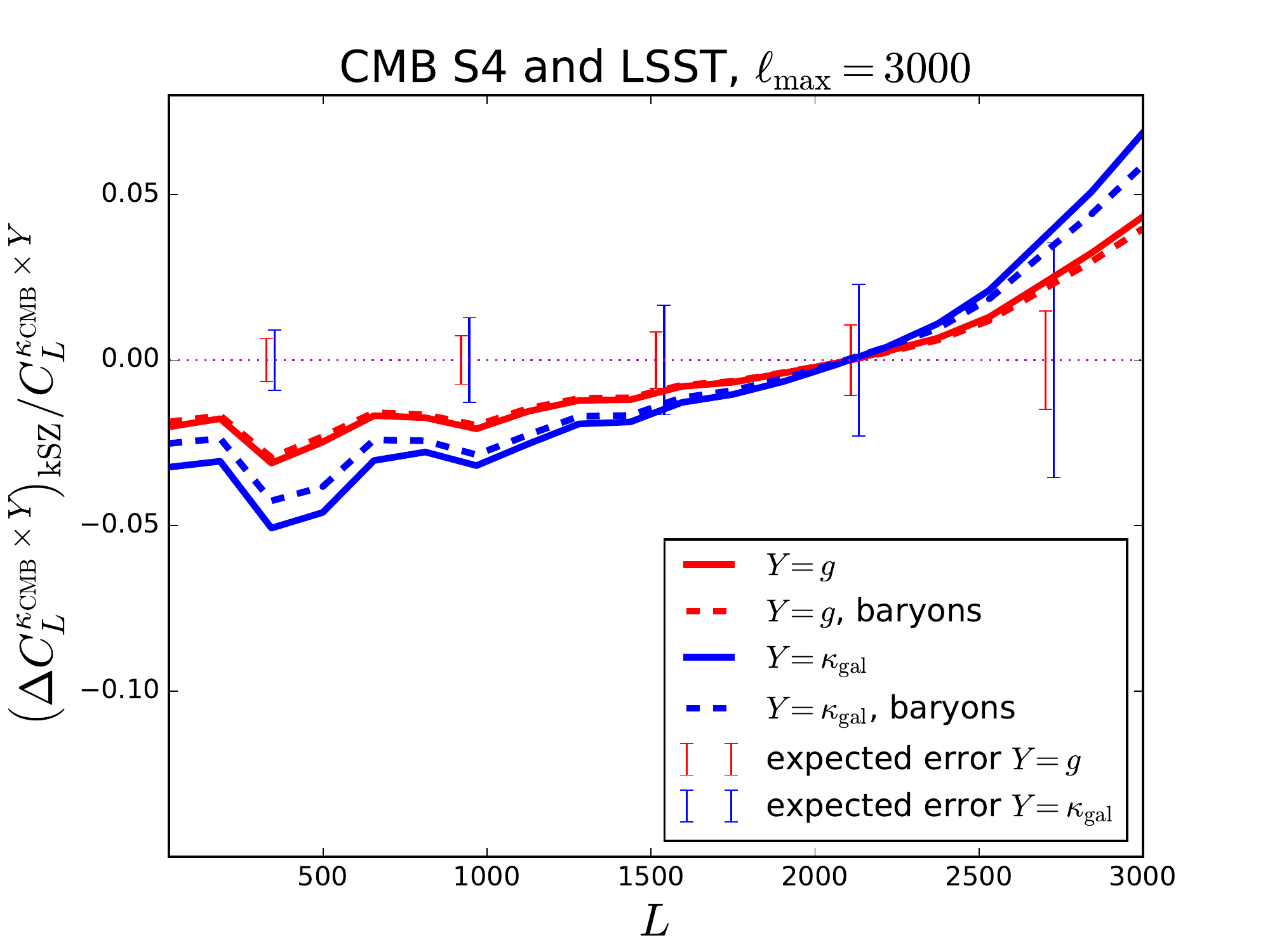} &   \includegraphics[width=8.5cm]{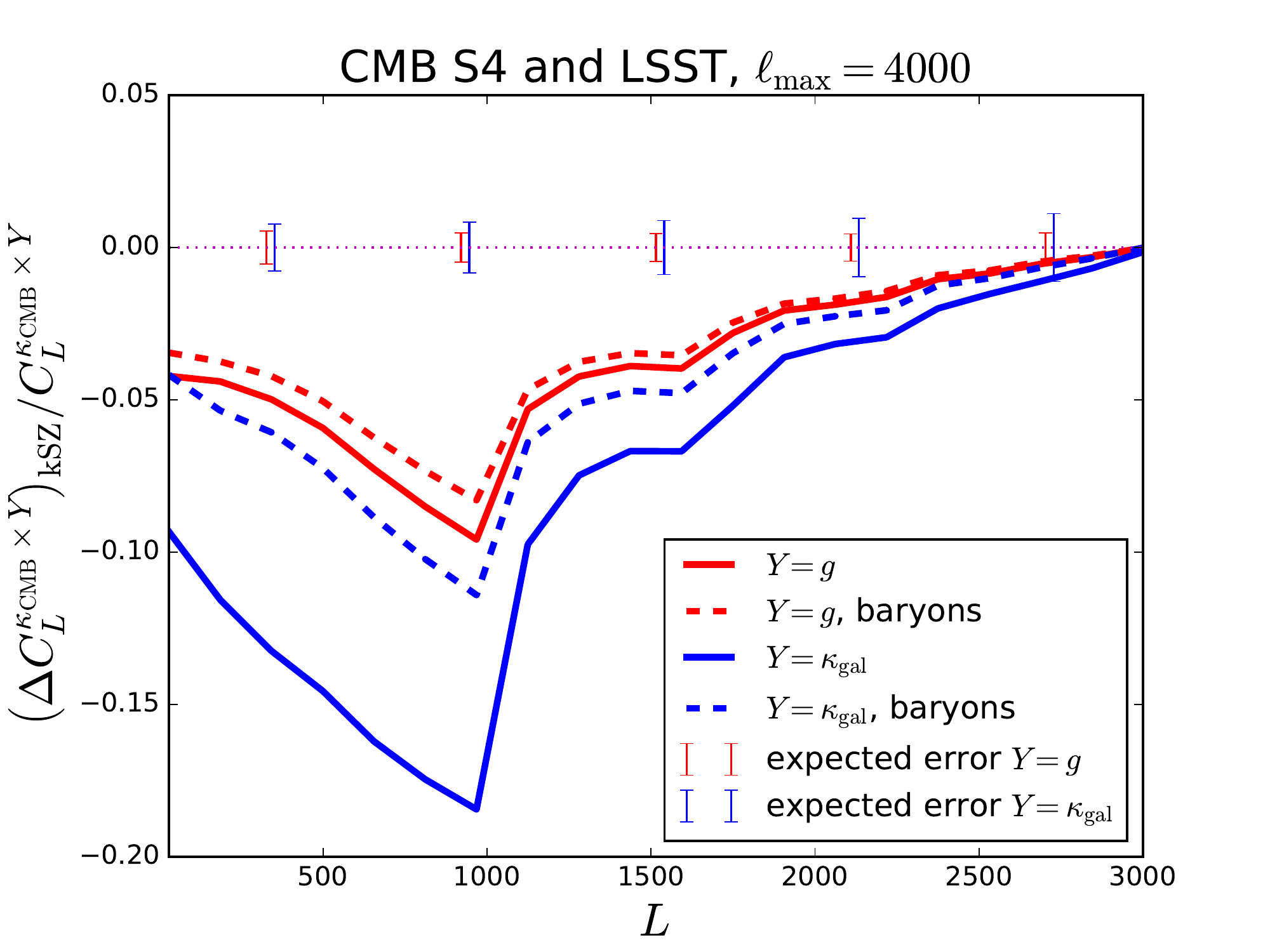} \\
\end{tabular}
\caption{Fractional bias to the cross-correlation of CMB lensing and galaxy density (denoted $Y=g$) or galaxy lensing ($Y = \kg$) maps from LSST.  From top to bottom, the panels show results for CMB lensing reconstructions from Planck SMICA, CMB-S3, and CMB-S4, respectively. The lensing reconstruction is performed on CMB temperature only, with $\ell_{\rm min} = 30$ and $\ell_{\rm max} = 3000$ (left panels) or $\ell_{\rm max} = 4000$ (right panels). The kSZ-induced bias is computed via Equation~\ref{eq:cross}.  The solid curves assume the gas perfectly traces the dark matter, while the dashed curves consider the effect of baryonic physics via the Jeans scale given in Equation~\ref{eq:kJ}. The bias is considerably larger than the forecasted error bars on such cross-correlation measurements for CMB-S3 and CMB-S4, shown here assuming $f_{\rm sky} = 0.44$ for the overlap between LSST and these surveys.} \
\label{fig:results_cross}
\end{figure}

Figure~\ref{fig:results_cross} shows the fractional bias to the cross-correlation of tracers (galaxies or galaxy lensing) with CMB lensing for the Planck, CMB-S3, and CMB-S4 configurations described above, i.e., $\left( \Delta C_L^{\kc \times g} \right)_{\rm kSZ}  / C_L^{\kc \times g}$. The top panel shows the results for CMB lensing convergence reconstructed from a Planck-like experiment, while the middle and bottom panels show the results for CMB-S3 and CMB-S4, respectively.  We include forecasted error bars computed from the standard analytic prescription including contributions from Gaussian sample variance and noise, with survey specifications as described in Sec.~\ref{subsec:configs} above, and assuming $f_{\rm sky} = 0.44$, in bins with $\Delta L \approx 600$.

Due to the relatively large noise level and beam size of Planck, the reconstruction technique mostly upweights large angular scales in the temperature map, which are the least affected by the kSZ contamination (since the primary CMB is significantly larger than the kSZ signal on these scales).  Thus, the kSZ-induced biases are generally within the statistical error bars for Planck.  As the noise level and beam size are lowered, smaller scales become important in the reconstruction, leading to a progressively larger bias due to the kSZ contamination.  For CMB-S3 and CMB-S4, the bias to the LSST galaxy lensing -- CMB lensing cross-correlation can be as large as $\approx 15$\%, when using reconstruction $\ell_{\rm max} = 4000$ and $\approx 5$\% for $\ell_{\rm max} = 3000$.  For comparison, the overall $S/N$ for the cross-correlation between LSST lensing and CMB-S4 lensing is expected to be $\approx 250$ and 160, respectively, using temperature reconstruction and the same values of $\ell_{\rm max}$.  Thus, the kSZ-induced bias is many times larger than the projected statistical errors on the cross-correlation.  The kSZ-induced bias thus requires careful treatment for efforts to calibrate the galaxy shear multiplicative bias via such cross-correlations~\cite{Vallinotto2012,Das2013,Liu2016,Baxter2016,2017arXiv170303383H,2016arXiv160701761S}, as well as for constraints on cosmology.

Moreover, Figure~\ref{fig:results_cross} shows that both the bias to the CMB lensing cross-correlation and the influence of baryonic physics are larger for LSST lensing than LSST galaxies.  This is because the kernel for galaxy lensing peaks at lower redshift than that for galaxy clustering, if the same LSST galaxies are used for both clustering and shape measurements.\footnote{Here, we assume that the same galaxy sample is used for clustering and as sources for lensing measurements. In this case, the lensing effect is produced by lower redshift structures, for which the kSZ signal and baryonic effects are larger.} The kSZ signal increases as redshift decreases, thus explaining the larger bias on lensing than galaxies, and the physical scale corresponding to a given angular scale is smaller at low redshift than high redshift, thus explaining the larger influence of baryons on lensing than galaxies (at fixed $L$).  If the galaxy sample is split into tomographic redshift bins, the lower redshift ones will be more affected by the kSZ bias and require better understanding of baryonic physics.

\subsection{Results: auto-power spectrum}
\label{subsec:results:auto}

\begin{figure}[ht]
\centering
\begin{tabular}{cc}
  \includegraphics[width=8.5cm]{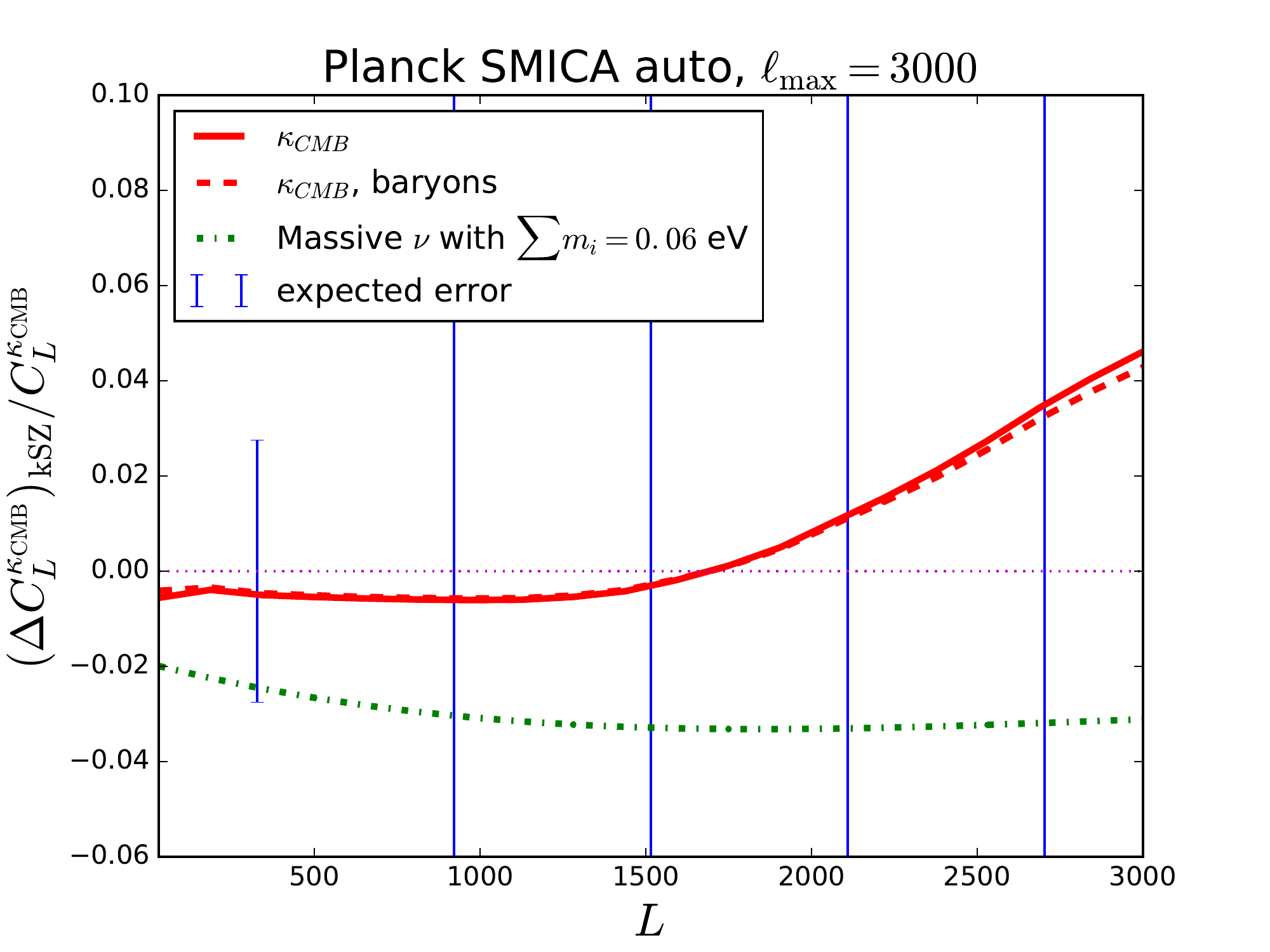} &   \includegraphics[width=8.5cm]{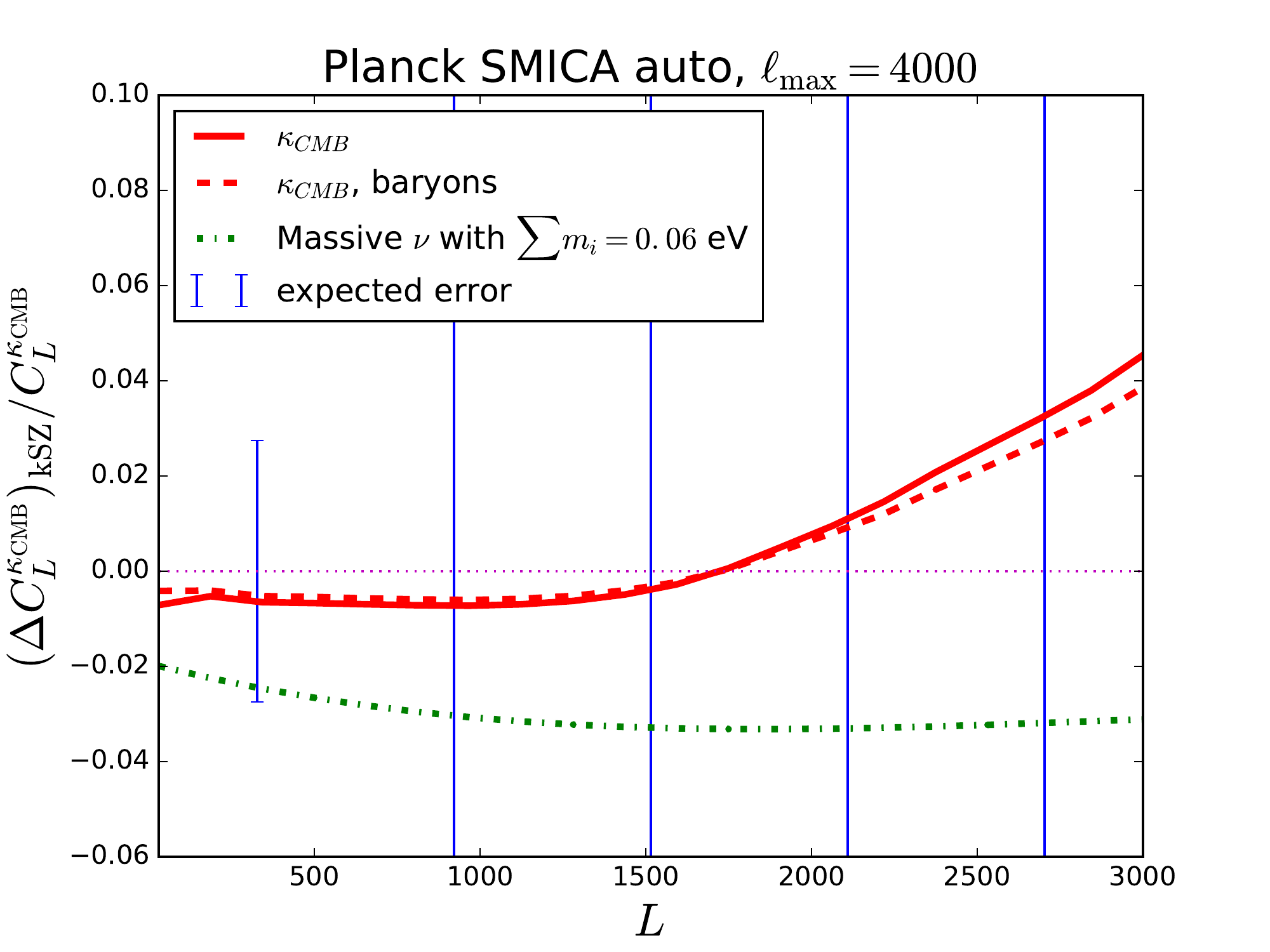} \\
  \includegraphics[width=8.5cm]{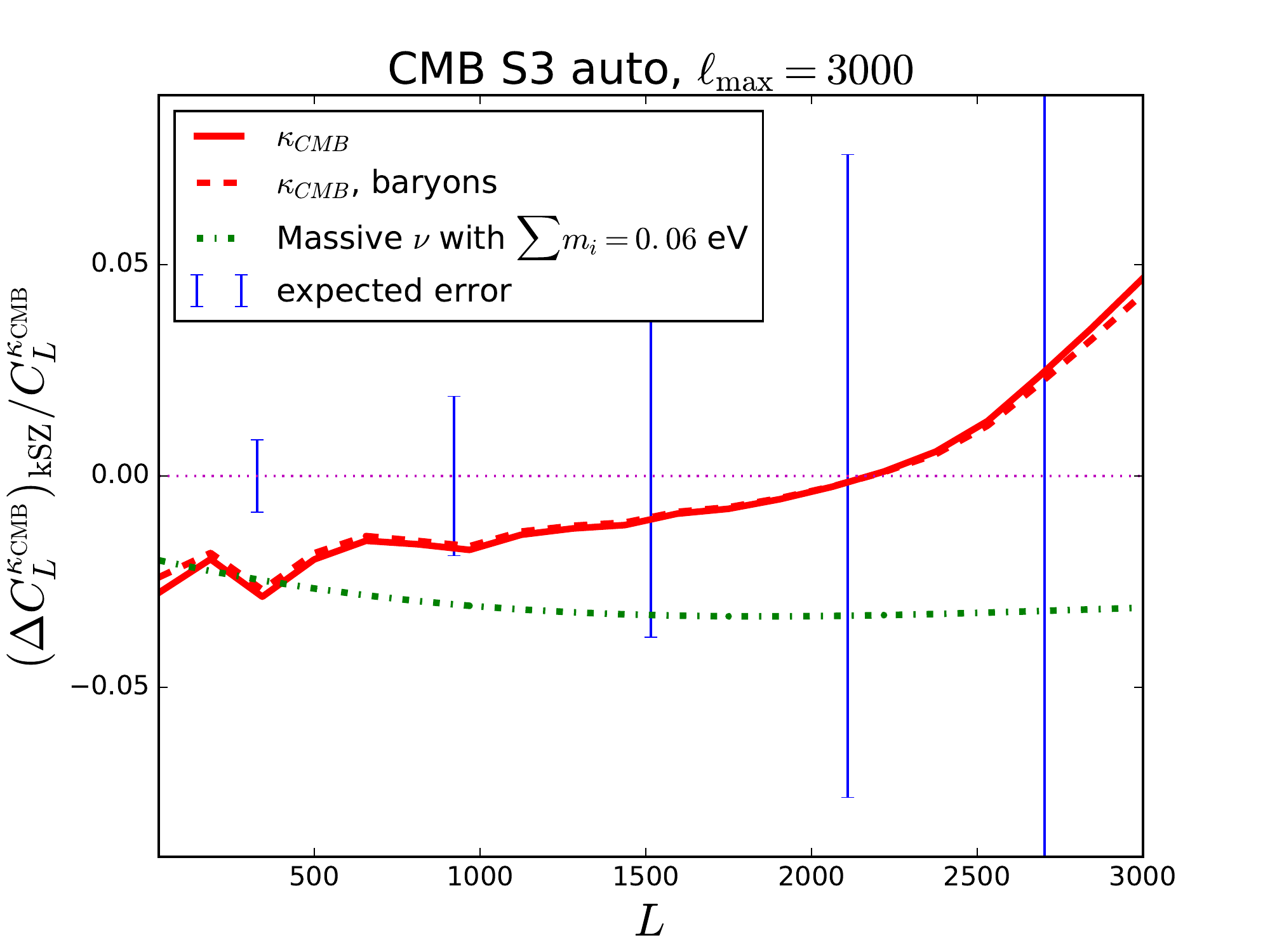} &   \includegraphics[width=8.5cm]{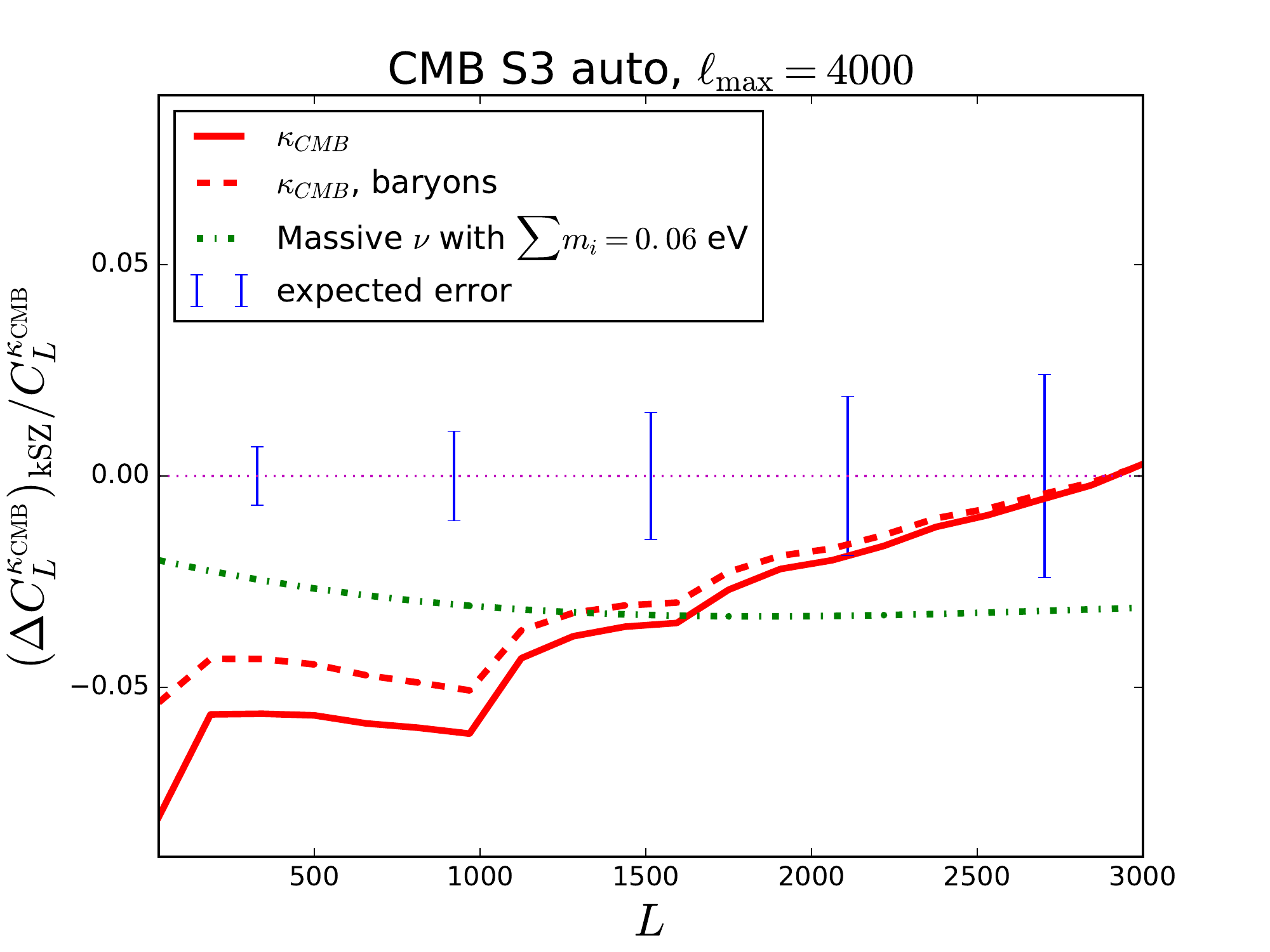} \\
  \includegraphics[width=8.5cm]{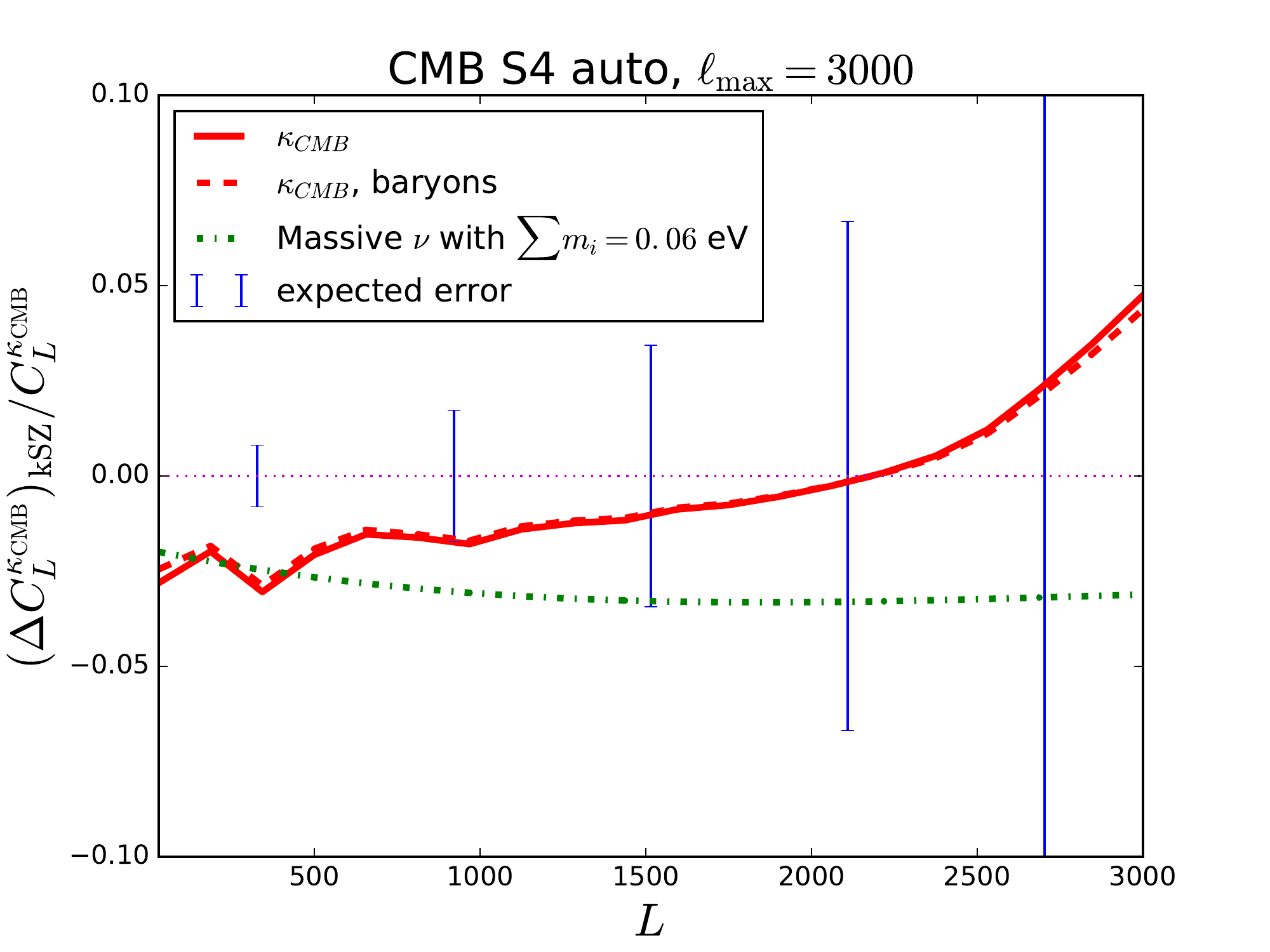} &   \includegraphics[width=8.5cm]{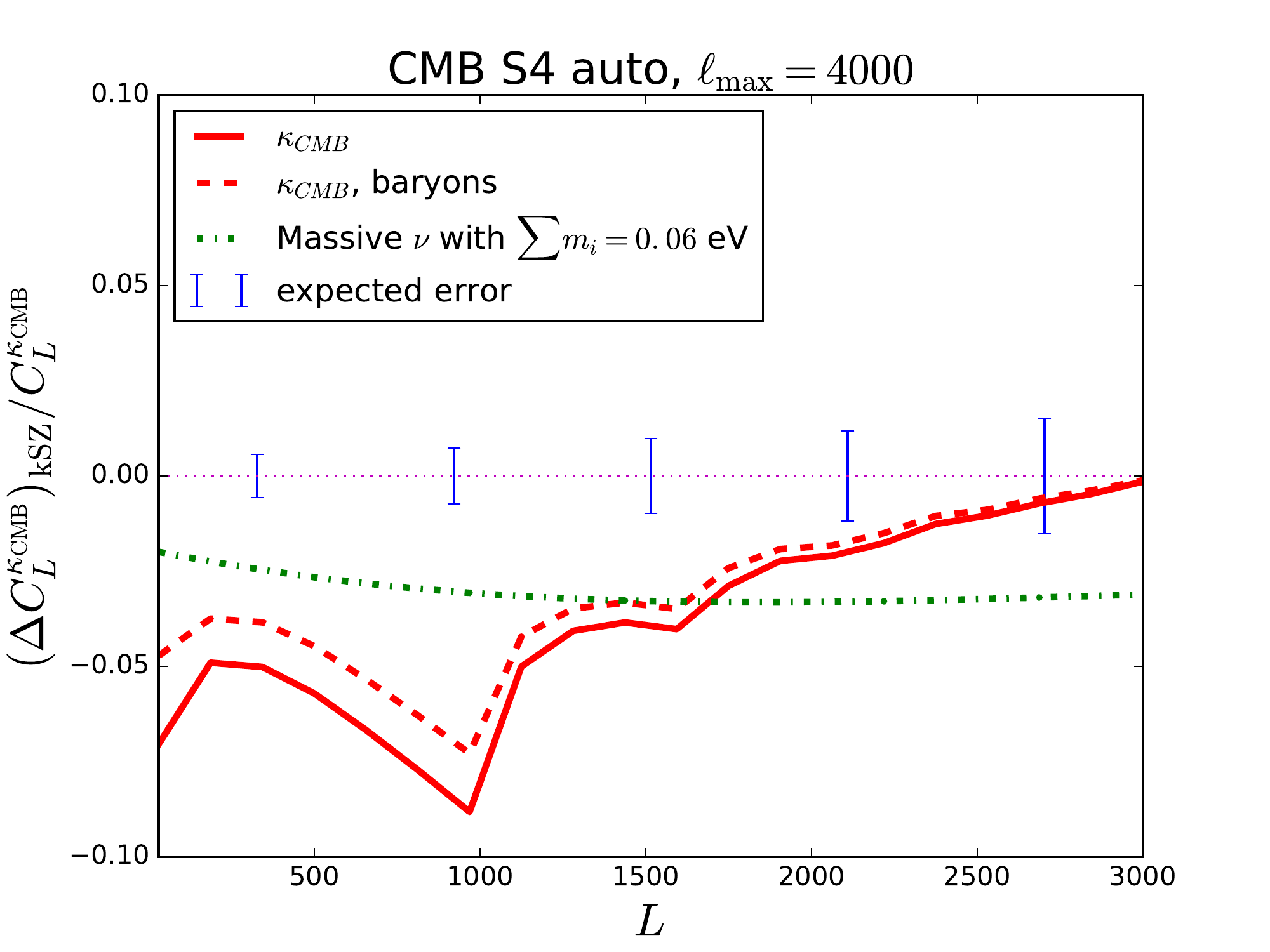} \\
\end{tabular}
\caption{Fractional bias to the CMB lensing auto-power spectrum arising from the term discussed in Section~\ref{sec:auto} (see Figure~\ref{fig:results_auto_full_sims} for an estimate of the full bias derived from simulations). The panel ordering is identical to Figure~\ref{fig:results_cross}. The lensing reconstruction is performed on temperature only, with $\ell_{\rm min} = 30$ and $\ell_{\rm max} = 3000$ (left panels) or $\ell_{\rm max} = 4000$ (right panels) and the kSZ-induced bias is computed via Equation~\ref{eq:auto}.  The solid curves assume the gas perfectly traces the dark matter, while the dashed curves consider the effect of baryonic physics via the Jeans scale given in Equation~\ref{eq:kJ}. For comparison, the dot-dashed curve shows the effect of massive neutrinos with a summed mass of 0.06 eV (the minimum mass in the normal hierarchy), which causes a $\approx 3$\% reduction in power compared to the fiducial case with massless neutrinos. The error bars show the statistical significance expected in 6 equally spaced multipole bins and assuming $f_{\rm sky} = 0.5$ for all experiments.}
\label{fig:results_auto}
\end{figure}

As discussed in Section~\ref{sec:auto}, the kSZ signal also leads to a bias on the CMB lensing auto-power spectrum. The dominant contribution to the bias is found by treating $\kappa_{\rm CMB}$ as a tracer of the low-redshift matter distribution, in analogy with the calculation in Section \ref{sec:cross}.  Other terms contribute to the auto-correlation as well, which are listed in Appendix~\ref{app:auto}.  We estimate the full bias from all terms in Section \ref{sec:fullauto}, modulo contributions from reionization.  At a minimum, the bias discussed in Section~\ref{sec:auto} should be present, which can be calculated in the analytic formalism presented earlier.  It is quantified in Figure~\ref{fig:results_auto} in terms of the fractional bias on $C_L^{\kc}$ for Planck SMICA, CMB-S3, and CMB-S4.  As in Figure~\ref{fig:results_cross}, we include forecasted error bars computed from the standard analytic prescription including contributions from Gaussian sample variance and noise, with survey specifications as described in Section ~\ref{subsec:configs}.  Note that our results in Section~\ref{sec:fullauto} confirm that the term computed analytically here is indeed the dominant term, so a comparison to the forecasted error bars is informative.

As in Figure~\ref{fig:results_cross}, the bias becomes larger when lowering the noise level and beam size, and simultaneously baryonic effects become more important. While the bias is sub-percent for the scales probed by the Planck satellite (and is thus smaller than the Planck statistical uncertainties on $C_L^{\kc}$), it can reach $\approx 5-10$\% for CMB-S3 or CMB-S4 when using $\ell_{\rm max} = 4000$ and about half of that for $\ell_{\rm max} = 3000$. If unaccounted for, it can lead to significant biases in cosmological parameters inferred from the CMB lensing auto-power spectrum.  As an example, we compute the fractional change in $C_L^{\kc}$ induced by massive neutrinos with a total mass of 0.06 eV, which is the minimum mass possible in the normal hierarchy, as compared to our fiducial model with massless neutrinos.  The suppression of the matter power spectrum below the neutrinos' free-streaming scale leads to a $\approx 3$\% suppression in $C_L^{\kc}$.  Detecting this effect is a major goal of upcoming CMB experiments, but as can be seen in Figure~\ref{fig:results_auto}, the kSZ-induced bias $\left( \Delta C_L^{\kc} \right)_{\rm kSZ}$ can be larger than the massive neutrino signal for both CMB-S3 and CMB-S4.  We will discuss possible mitigation strategies to overcome the kSZ-induced bias in Section~\ref{sec:mitigation}.

\section{Numerical Simulations}
\label{sec:sims}
\subsection{Validation of Analytic Formalism}
\label{sec:validation}
We validate the analytic formalism presented above by comparing to results measured from the cosmological simulation described in Ref.~\cite{2010ApJ...709..920S}.  In this analysis, a lightcone was extracted from a large dark matter $N$-body simulation ($1$~Gpc$/h$ on a side).  It was then post-processed with a variety of baryonic physics prescriptions to generate realistic simulations of several signals in the microwave sky (thermal SZ, kSZ, cosmic infrared background, radio point sources, and Galactic thermal dust).  The kSZ power spectrum extracted from this simulation is consistent with upper limits from ACT~\cite{Sievers2013} and SPT~\cite{George2015}.  However, we note that a high-pass filter was applied to the kSZ field in this simulation in order to correct an overprediction of the intergalactic medium kSZ signal on large angular scales (which resulted from the lightcone construction).  In particular, a filter $w(\ell) = 1 - e^{-(\ell/500)^2}$ was applied to the kSZ map to suppress the large-scale excess.  Thus, the kSZ field at $\ell \lesssim 1000$ may not be expected to match analytic calculations perfectly, which should be kept in mind when comparing the simulation and analytic results below.

For our purposes, the most important feature of this simulation is that the extragalactic signals are all realistically correlated with one another.  In particular, a $\kappa_{\rm CMB}$ map was generated by summing the mass in redshift shells extracted from the $N$-body volume using the CMB lensing kernel in Equation~\ref{eq.WkappaCMBdef} (no ray-tracing was performed, i.e., the Born approximation was assumed), and the kSZ signal was generated self-consistently from ionized gas ``pasted'' onto the same large-scale structure realization.  The simulated maps cover an octant of the sky, which was replicated eight times to yield full-sky HEALPix maps.  The kSZ map is provided at resolution $N_{\rm side} = 8192$, while the $\kc$ map is provided at $N_{\rm side}=4096$.  For consistency and computational efficiency, we downgrade the kSZ map to $N_{\rm side}=4096$ (accounting appropriately for the pixel window function) and work at this resolution throughout the following analysis.

We use only the $\kappa_{\rm CMB}$ and kSZ maps from the simulation.  In particular, due to a sign error in the deflection calculation used to lens the primary CMB in the simulated temperature maps, we do not use the provided lensed (or unlensed) CMB temperature maps.  We instead generate an unlensed CMB temperature map from a CMB power spectrum computed with {\tt camb},\footnote{\url{http://camb.info}} with cosmological parameters matching those used in the simulation (the temperature map is generated at $N_{\rm side} = 4096$ from a CMB power spectrum extending to $\ell =  10000$).  We then use {\tt LensPix}\footnote{\url{http://cosmologist.info/lenspix/}} to lens this CMB map with the deflection field computed from the simulated $\kappa_{\rm CMB}$ map.\footnote{We verify that the power spectrum of the $N$-body simulation-derived $\kc$ map matches the non-linear prediction from {\tt camb} (which uses Halofit~\cite{2003MNRAS.341.1311S,2012ApJ...761..152T}) very accurately to $L=5000$, which is more than sufficient for our purposes.}  We verify that the power spectrum of the resulting lensed CMB temperature map matches the {\tt camb} prediction to effectively exact precision up to $\ell=5000$, which is higher than any of the $\ell_{\rm max}$ values used in the reconstructions in this paper.

We use the full-sky CMB lensing reconstruction algorithm provided in {\tt LensPix} to reconstruct maps of $\hat{\kappa}_{\rm CMB}$.  We consider each of the three experimental configurations given in Table~\ref{tab:CMBconfig}, which define the properties of the filter functions used in the estimator.  We use the same multipole range for the reconstructions as considered earlier: $\ell_{\rm min} = 30$ and $\ell_{\rm max} = 3000$ or 4000.  Note that the kSZ-related biases can be mitigated to some extent by decreasing $\ell_{\rm max}$ at the cost of decreased $S/N$ on the CMB lensing reconstruction, as shown in Figures~\ref{fig:results_cross} and~\ref{fig:results_auto} and discussed further in Section~\ref{sec:mitigation}.   In this subsection, we only compute cross-power spectra of reconstructed maps with input maps, and thus we avoid the $N^{(0)}$ and $N^{(1)}$ biases that afflict $\hat{\kappa}_{\rm CMB}$ auto-power spectra.  In the following subsection, we will construct Gaussian simulations that by definition have identical $N^{(0)}$ and $N^{(1)}$ biases as the ``true'' simulation.  We mitigate (most of) the $N^{(2)}$ bias by following the standard practice of using lensed CMB power spectra in the reconstruction filters~\cite{Hanson2011}.  Note that the filter denominator explicitly includes the kSZ power (in addition to lensed CMB and noise), as would be the case in an actual data analysis.  We do not explicitly add noise to any of the simulated maps, but we verify that this does not bias any of the power spectrum results presented here.

We first verify that the cross-power spectrum of the reconstructed convergence ($\hat{\kappa}_{\rm CMB}$) with the true convergence ($\kc$), $\langle \hat{\kappa}_{\rm CMB} \kappa_{\rm CMB} \rangle$, matches theoretical expectations.  For Planck SMICA, $\langle \hat{\kappa}_{\rm CMB} \kappa_{\rm CMB} \rangle$ matches $\langle \kc \kc \rangle$ well.  For CMB-S3 and CMB-S4, a small residual difference is present (a fractional deficit $\approx 5 \%$ on large scales for $\ell_{\rm max} = 4000$ and roughly half this value for $\ell_{\rm max} = 3000$).  However, this residual bias is consistent with estimates of the additional $N^{(2)}$ bias arising from the sub-optimal choice of filter weights~\cite{2011JCAP...03..018L,2017PhRvD..95d3508P} and the $N^{(3/2)}$ bias due to the non-zero lensing potential bispectrum (our map includes only the non-linear growth contributions to this bias, and not the post-Born contributions)~\cite{2016PhRvD..94d3519B}.  Regarding the residual $N^{(2)}$ bias, as discussed in~\cite{2017PhRvD..95d3508P}, an unbiased temperature-reconstructed $\hat{\kappa}_{\rm CMB}$ power spectrum at CMB-S4 noise levels requires the use of the non-perturbative gradient power spectrum $C_{\ell}^{\tilde{\Theta} \nabla \tilde{\Theta}}$ in the filter weights, rather than $C_{\ell}^{\tilde{\Theta} \tilde{\Theta}}$.  As our focus is not on the reconstructed convergence auto-power spectrum, $\langle \hat{\kappa}_{\rm CMB} \hat{\kappa}_{\rm CMB} \rangle$, but rather on the kSZ-related biases, we do not consider these higher-order $\hat{\kappa}_{\rm CMB}$ biases further, and proceed to use the estimator as described above.

We proceed to validate the analytic theory presented earlier by measuring the kSZ-induced bias arising from the correlation of the kSZ field with the $\kc$ field (Term B in the terminology of Appendix~\ref{app:auto}).  We run the lensing reconstruction estimator on the simulated kSZ map to obtain a map of $\hat{\kappa}_{\rm CMB}^{\rm kSZ}$, which we cross-correlate with the (true) input $\kc$ field, $\langle \hat{\kappa}_{\rm CMB}^{\rm kSZ} \kc \rangle$.  Error bars are obtained from the standard analytic formula for cross-correlations with $f_{\rm sky} = 0.125$, assuming the Gaussian approximation (which is valid due to the wide multipole bins considered, with $\Delta \ell = 200$).  The error bars are small as there is no noise added to the maps.  We assess the kSZ-induced bias by comparing $\langle \hat{\kappa}_{\rm CMB}^{\rm kSZ} \kc \rangle$ to $\langle \kc \kc \rangle$.

To compare the cross-correlation formalism between the simulations and analytic theory, we extract a sample of tracer halos from the catalogs provided by Ref.~\cite{2010ApJ...709..920S}, and measure cross-correlations of this sample with the input and reconstructed convergence maps.  The sample is defined by selecting all halos with redshift $0.1 < z < 0.8$ and halo mass $M_{200c} > 5 \times 10^{13} \, M_{\odot}$, yielding 131388 objects.  Using the sky position of each halo, we generate a map of the tracer number density fluctuation, $\delta_g({\bf \hat{n}}) = (n_g({\bf \hat{n}}) - \bar{n}_g)/\bar{n}_g$.  This map is only defined on the original simulation octant, and thus we must apply a mask defining this octant when using the $\delta_g$ map in cross-correlations.  We apodize the mask using a Gaussian taper with FWHM = 30 arcmin.  We correct for its effect in the power spectrum results using a simple $f_{\rm sky}$ factor, which is sufficiently accurate given the mask's simple structure and large sky fraction, as well as the wide multipole bins considered.  We cross-correlate this map with the (true) input convergence field $\kc$, as well as with the $N$-body and kSZ reconstructions described above.  From the measurement of $\langle \kc \delta_g \rangle$, we determine the linear bias of the tracer sample, which is needed for the analytic calculation in Equation~\ref{eq:cross}.  We verify that $\langle \hat{\kappa}_{\rm CMB} \delta_g \rangle$ agrees well with $\langle \kc \delta_g \rangle$ (up to a small residual bias consistent with $N^{(3/2)}$~\cite{2016PhRvD..94d3519B}), and thus assess the kSZ-induced bias by comparing $\langle \hat{\kappa}_{\rm CMB}^{\rm kSZ} \delta_g \rangle$ to $\langle \kc \delta_g \rangle$.

To approximately model the effects of star formation, feedback, and Helium reionization which are present in the simulations, we compare the amplitude of the kSZ power spectrum from the simulations to that expected in our theoretical framework, defining an ``effective'' $f_{\rm free}^{\rm eff}$ as $(C_{\ell}^{\rm kSZ})_{\rm simulations} = (f_{\rm free}^{\rm eff})^2 (C_{\ell}^{\rm kSZ})_{\rm theory}$.\footnote{Here $(C_{\ell}^{\rm kSZ})_{\rm theory}$ is calculated with $f_{\rm free} = 1$.}   By fitting the simulations over the range $\ell = 1500 - 3000$, we find $f_{\rm free}^{\rm eff} \approx 0.7$, which is the value that we use in the comparison between the analytic formalism and simulation results below.  Note that this is analogous to measuring the amplitude of the kSZ signal in the real Universe, and then using the measured amplitude to predict the kSZ-induced bias to CMB lensing.

\begin{figure}[t]
\includegraphics[width=10cm]{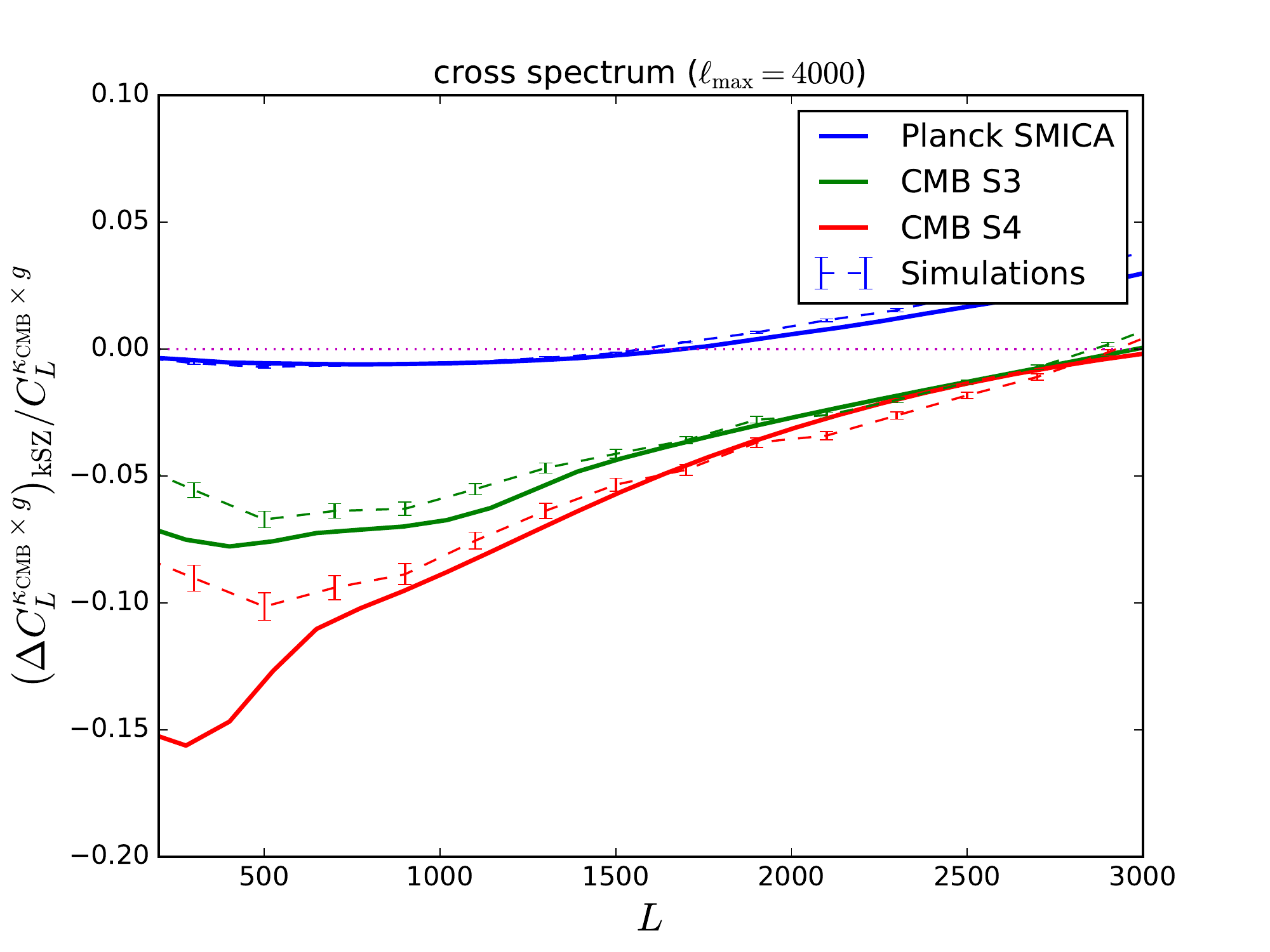}
\caption{Comparison of simulation-derived (dashed curves with error bars) and analytic predictions (solid curves) for the fractional kSZ-induced bias on the cross-correlation of CMB lensing and tracers, for various experimental configurations.  All reconstructions here use $\ell_{\rm max} = 4000$.  The agreement between the theory and simulations is generally good.  The small discrepancies seen on large scales for the CMB-S4 case may be related to filtering applied to the kSZ field in the simulation or differing treatments of baryonic effects between the analytic calculation and the simulation. Note that the cosmological parameters and $f_{\rm free}$ here differ from the fiducial ones in the rest of the paper and are chosen to match those in the simulations.}
\label{fig:comparison_sims}
\end{figure}

Figure~\ref{fig:comparison_sims} presents a comparison of the tracer cross-correlation results from the simulations to those derived from the analytic formalism described earlier, including the baryon effects quantified through the filtering scale (c.f. Equation~\ref{eq:cross}).  The analytic calculation uses identical cosmological parameters to those used in the numerical simulation, as well as a tracer bias matching that for the sample extracted from the simulation.  The agreement for the fractional bias on $C_L^{\kc \times g}$ is generally good, although minor discrepancies are seen for the CMB-S4 case at low multipoles.  It is possible that this is related to the large-scale filtering applied to the kSZ field in the simulation (see discussion earlier) or to differences between the baryonic effects in the simulation and those in the analytic calculation.

\begin{figure}[t]
\includegraphics[width=10cm]{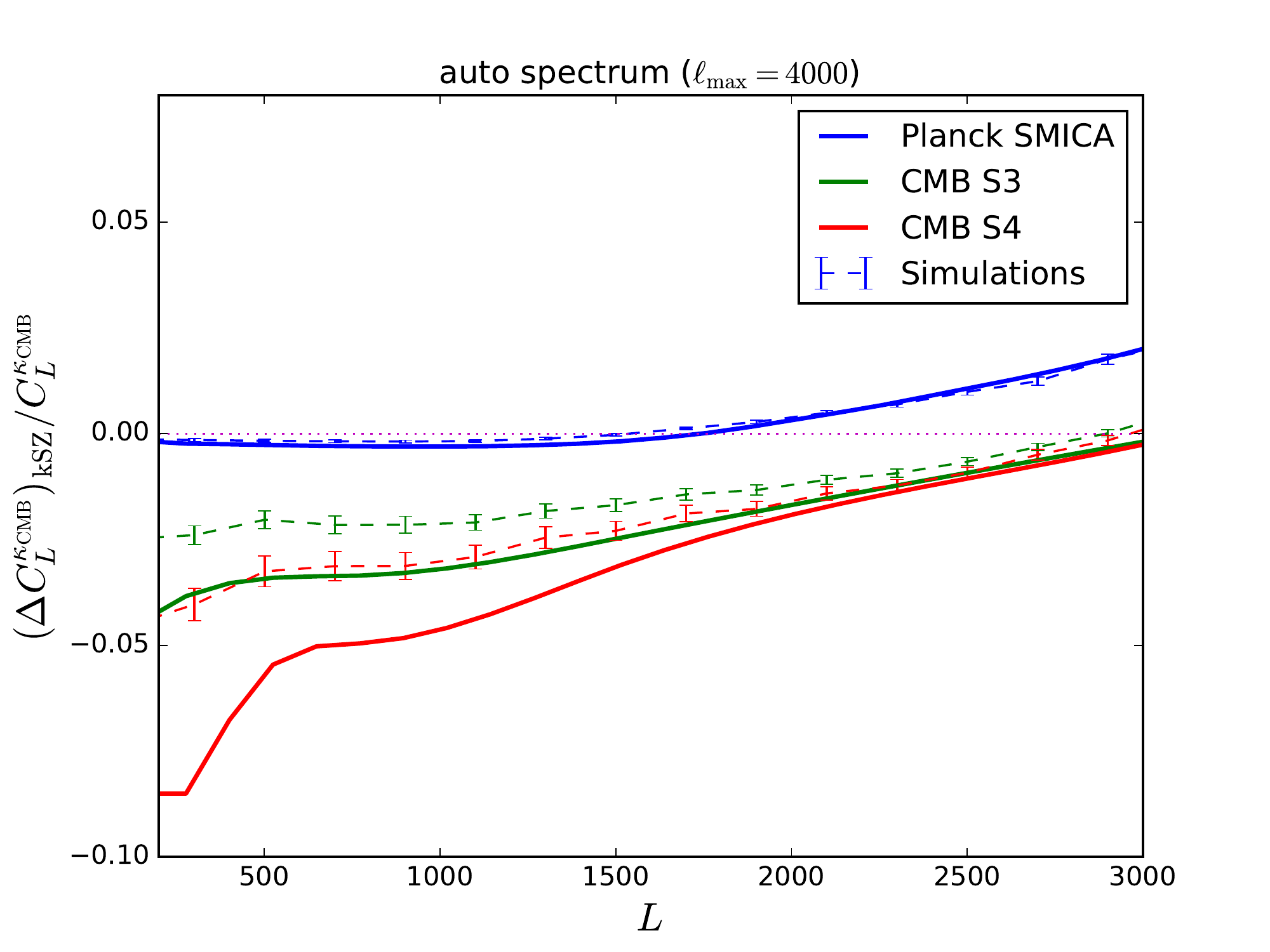}
\caption{Comparison of simulation-derived (dashed curves with error bars) and analytic predictions (solid curves) for the fractional kSZ-induced bias on the CMB lensing auto-power spectrum, for various experimental configurations, including only the term discussed in Section~\ref{sec:auto} (see Figure~\ref{fig:results_auto_full_sims} for an estimate of the full bias including all terms).  All reconstructions here use $\ell_{\rm max} = 4000$.  The agreement between the theory and simulations is generally good, though somewhat less precise than in Figure~\ref{fig:comparison_sims}.  The discrepancies seen on large scales for the CMB-S3 and CMB-S4 cases may be due to filtering applied to the kSZ field in the simulation or differing treatments of baryonic effects (particularly He reionization) between the analytic calculation and the simulation. Note that the cosmological parameters and $f_{\rm free}$ here differ from the fiducial ones in the rest of the paper and are chosen to match those in the simulations.}
\label{fig:comparison_sims_auto}
\end{figure}

Figure~\ref{fig:comparison_sims_auto} shows a comparison of the auto-power spectrum results from the simulations to those obtained from the analytic formalism described earlier, specifically the term given in Equation~\ref{eq:auto}.  The agreement for the fractional bias on $C_L^{\kc}$ is again reasonable, although the theory calculation appears to overpredict the bias on large scales for the CMB-S3 and CMB-S4 cases.  We again suspect that these issues could be related to the filtering of the simulation kSZ map or to baryonic effects.  Also, at $z \gtrsim 3$, Helium is only partly ionized and therefore the number of free electrons is lower at those redshifts.  This only affects the bias to the CMB lensing auto-power spectrum (and is present in the simulations), and not the bias on cross-correlations with tracers at lower redshift. The $f_{\rm free}^{\rm eff}$ determination discussed above partially captures this effect, but is not exact due to the differing redshift kernels of the kSZ power spectrum (used to determine $f_{\rm free}^{\rm eff}$) and the kSZ-induced bias on the $\kc$ power spectrum. Modeling the effects of He reionization is beyond the scope of this paper and is left to future work.

Overall, the agreement in Figures~\ref{fig:comparison_sims} and~\ref{fig:comparison_sims_auto} provides a strong check of our analytic formalism and verifies that the magnitude of the kSZ-induced bias cannot be neglected for CMB-S3 or CMB-S4 lensing measurements.

\subsection{Estimate of Full Auto-Spectrum Bias}
\label{sec:fullauto}
We proceed to use the simulation maps to estimate the full bias to the reconstructed CMB lensing auto-power spectrum arising from the kSZ effect, modulo contributions from reionization, which are not included in the simulation.  The tracer cross-correlation bias is fully described by Equation~\ref{eq:cross}, and thus the analytic formalism captures the full effect.  The auto-correlation bias includes the contribution from Equation~\ref{eq:auto} (Term B in~Appendix~\ref{app:auto}), which our analytic formalism describes, but also includes contributions from two additional terms that are more difficult to compute analytically.  The first, labeled Term C in~Appendix~\ref{app:auto}, is simply the alternative Wick contraction of the four fields present in Term B (such terms were labeled ``secondary contractions'' in~Ref.~\cite{2014JCAP...03..024O}).  The second, labeled Term E in~Appendix~\ref{app:auto}, arises from the non-zero connected trispectrum of the kSZ signal.

Instead of breaking the bias down into its constituent terms, we estimate the sum of all three (Terms B+C+E) via the following procedure.  First, we generate ten Gaussian kSZ realizations ($\Theta^{\rm kSZ,g}$) with a power spectrum precisely matching that of the true kSZ map ($\Theta^{\rm kSZ,sim}$) described in the previous subsection.  Second, we add each of these Gaussian kSZ maps to the lensed CMB temperature map ($\tilde{\Theta}^{\rm sim}$) described above: $\Theta^{\rm tot,g} = \Theta^{\rm kSZ,g} + \tilde{\Theta}^{\rm sim}$.  We also add the true kSZ map to the lensed temperature map: $\Theta^{\rm tot} = \Theta^{\rm kSZ,sim} + \tilde{\Theta}^{\rm sim}$.  We then run the {\tt LensPix} reconstruction algorithm on the ten realizations of $\Theta^{\rm tot,g}$ (obtaining maps of $\hat{\kappa}_{\rm CMB}^{\rm tot,g}$) and the map containing the true kSZ field $\Theta^{\rm tot}$ (obtaining maps of $\hat{\kappa}_{\rm CMB}^{\rm tot}$).  By construction, the biases on the auto-power spectrum of the reconstructed lensing fields for all of these maps are identical, except for terms involving mixtures of the non-Gaussian $\kc$ and kSZ fields (or the non-Gaussian kSZ field alone), which are the biases we want to estimate.  Thus, we can measure the kSZ-induced biases by simply subtracting the reconstruction auto-power spectra
\be
\left( \Delta C_L^{\kc} \right)_{\rm kSZ} = C_L^{\hat{\kappa}_{\rm CMB}^{\rm tot}} - \langle C_L^{\hat{\kappa}_{\rm CMB}^{\rm tot,g}} \rangle_{\rm avg},
\ee
where the angle brackets in the second term indicate an average over the ten Gaussian kSZ realizations.\footnote{Note that the residual $N^{(2)}$ and $N^{(3/2)}$ biases discussed earlier will cancel in this procedure, in addition to $N^{(0)}$ and $N^{(1)}$.}  We estimate error bars on $\left( \Delta C_L^{\kc} \right)_{\rm kSZ}$ from the scatter amongst the ten realizations.  We note that this calculation is the first full estimate in the literature (to our knowledge) of all contributions (i.e., Terms B+C+E) to a secondary-induced CMB lensing auto-power spectrum bias.

\begin{figure}[ht]
\centering
\begin{tabular}{cc}
  \includegraphics[width=9cm]{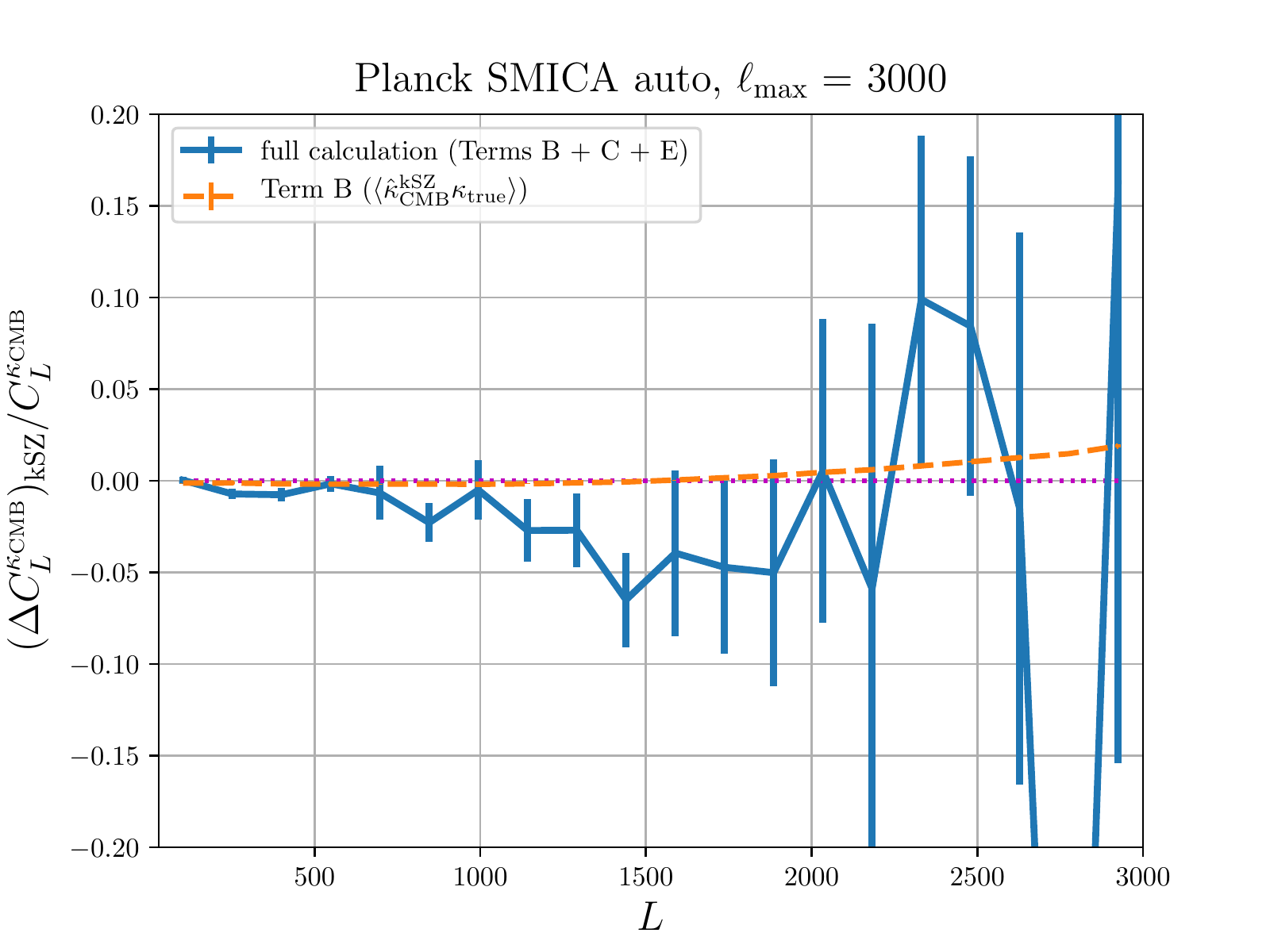} &   \includegraphics[width=9cm]{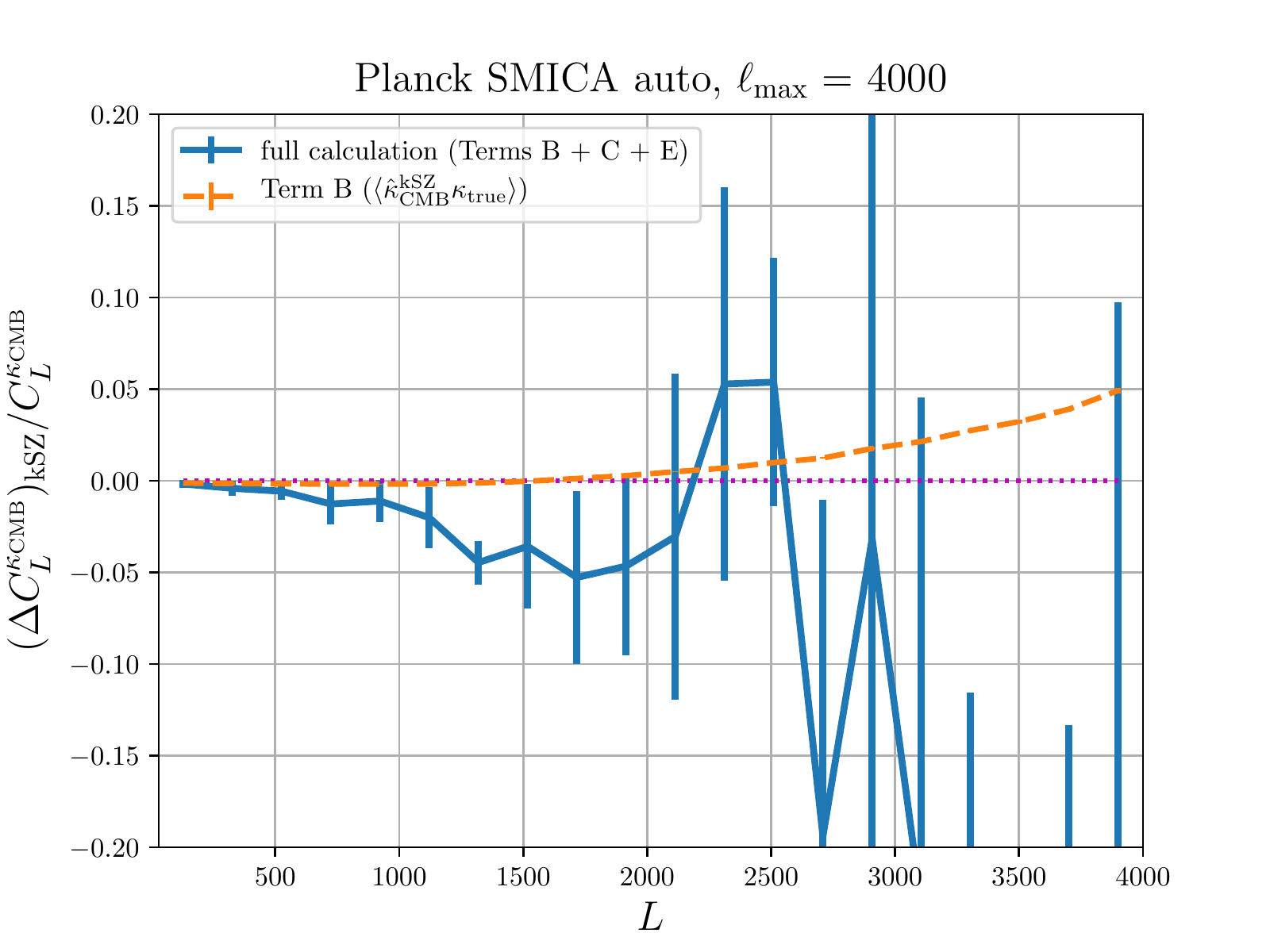} \\
  \includegraphics[width=9cm]{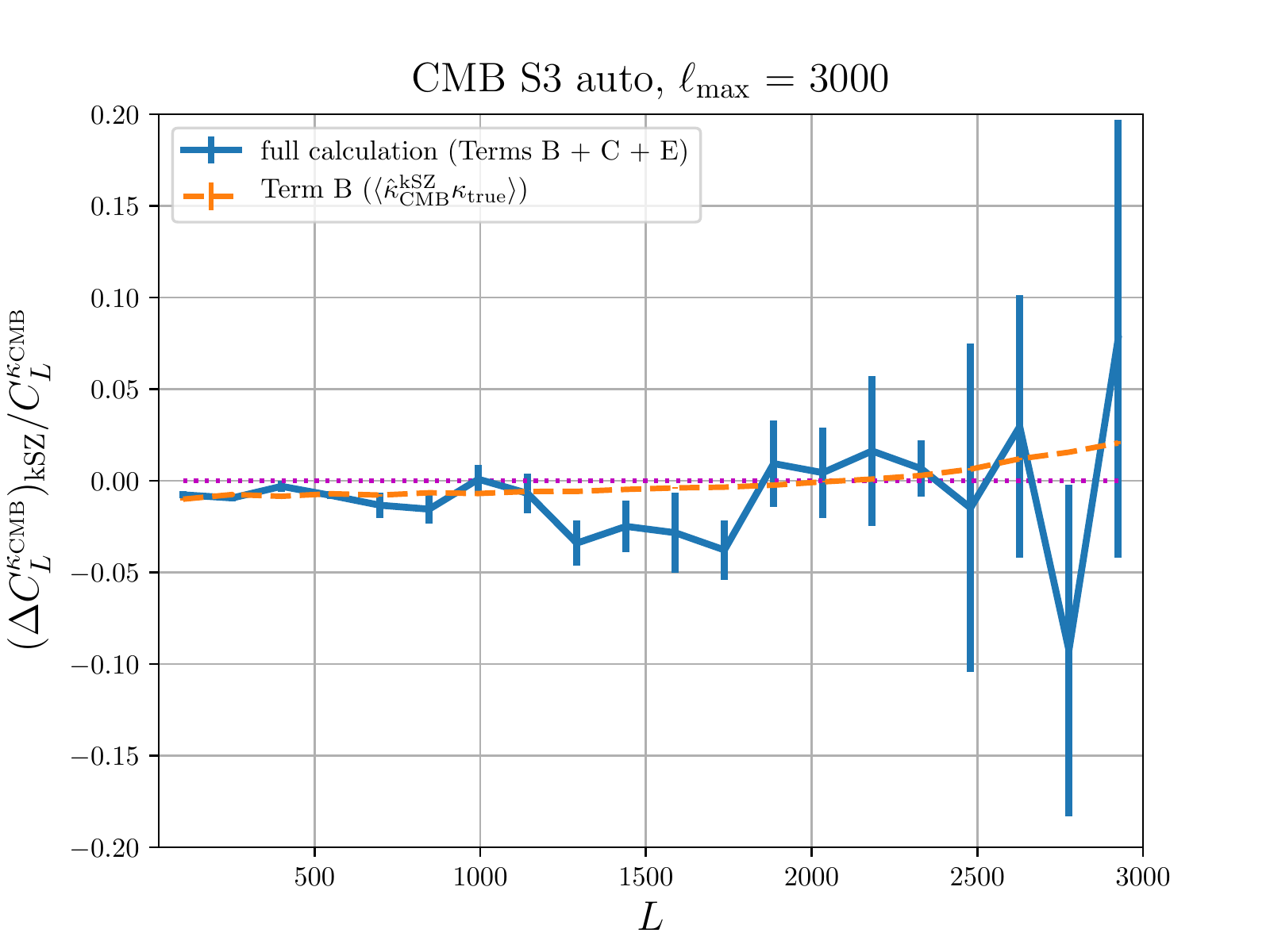} &   \includegraphics[width=9cm]{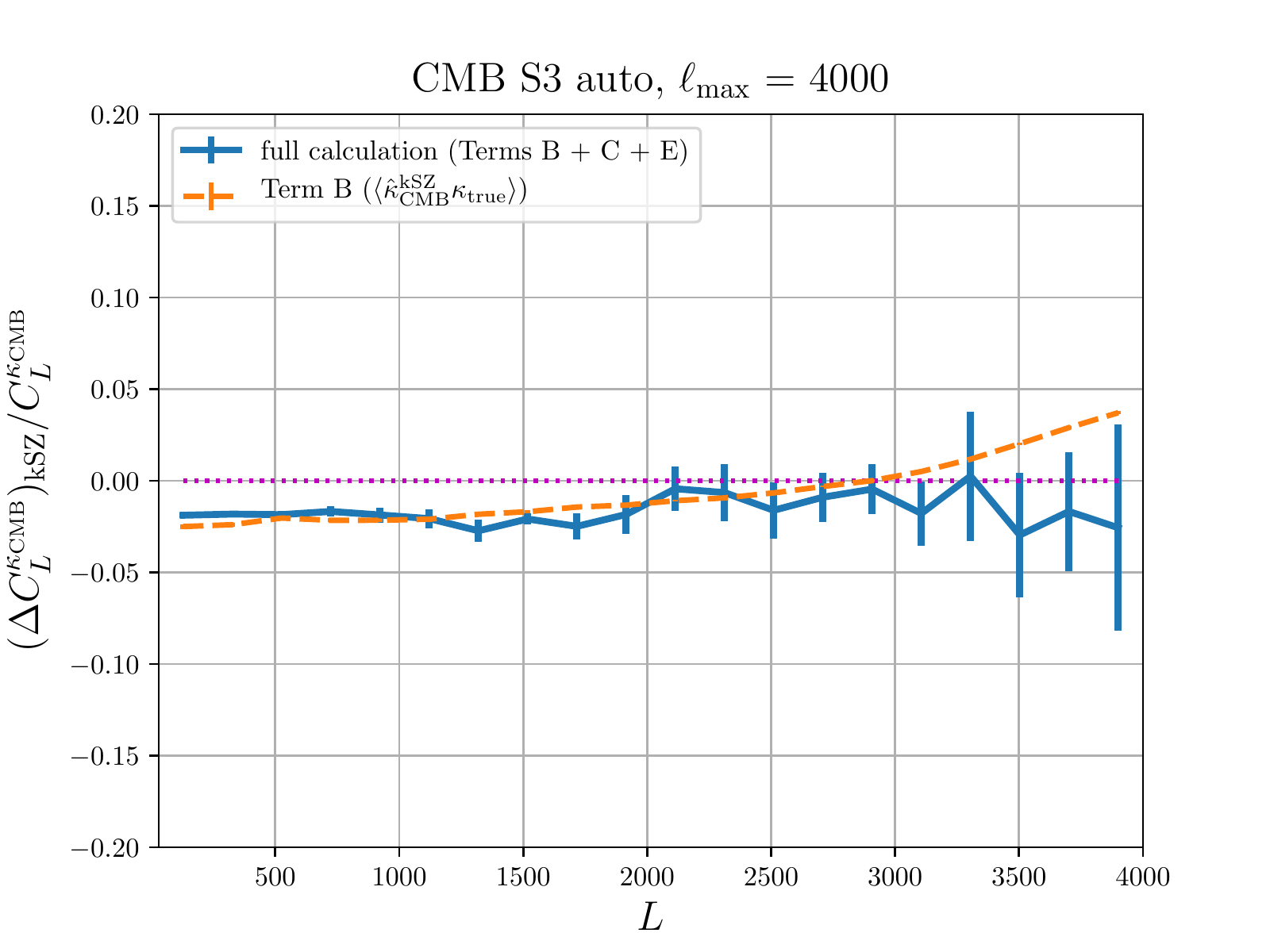} \\
  \includegraphics[width=9cm]{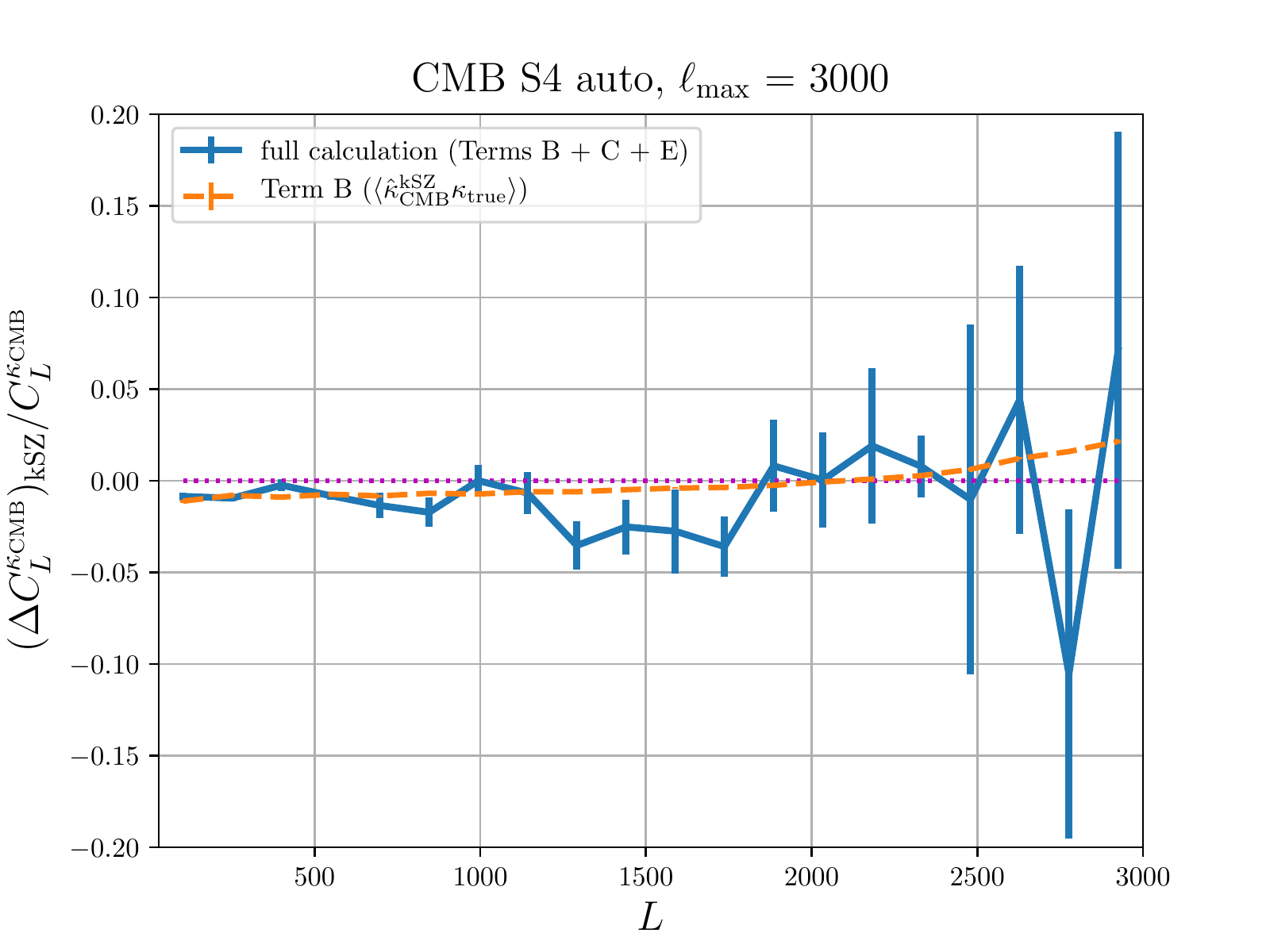} &   \includegraphics[width=9cm]{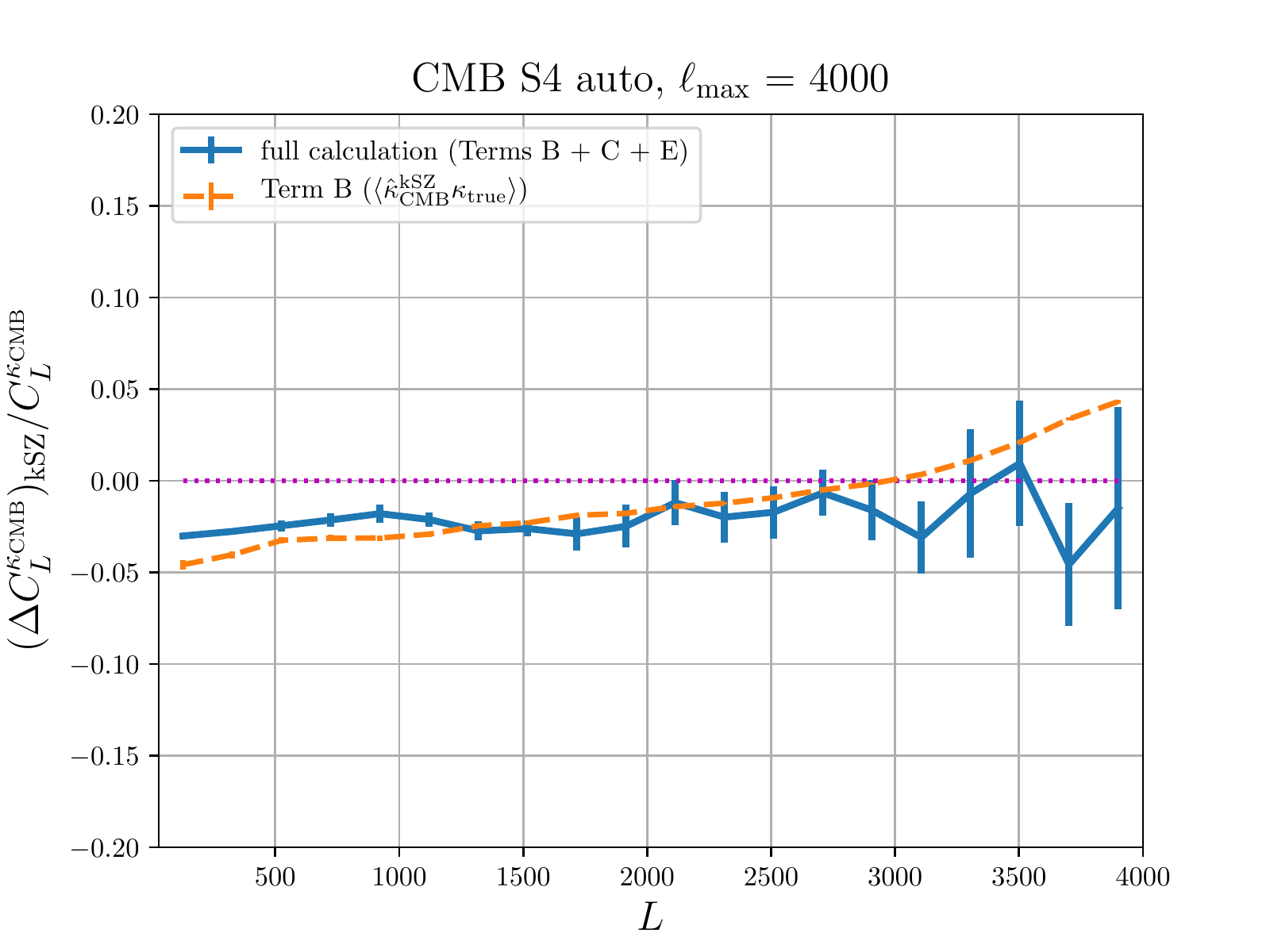} \\
\end{tabular}
\caption{Fractional bias to the reconstructed CMB lensing auto-power spectrum arising from the sum of all terms discussed in Appendix~\ref{app:auto} (Terms B+C+E), as estimated via the procedure described in Sec.~\ref{sec:fullauto}. The panel ordering is identical to Figures~\ref{fig:results_cross} and~\ref{fig:results_auto}. The lensing reconstruction is performed on temperature only, with $\ell_{\rm min} = 30$ and $\ell_{\rm max} = 3000$ (left panels) or $\ell_{\rm max} = 4000$ (right panels) and the kSZ-induced bias is computed by comparing reconstructions involving Gaussian kSZ maps to those using the true non-Gaussian kSZ map (see Sec.~\ref{sec:fullauto}).  The solid curves show our estimate of the full kSZ-induced bias (modulo reionization contributions), while the dashed curves show the contribution of Term B (i.e., Equation~\ref{eq:auto}) only, as plotted in Figure~\ref{fig:comparison_sims_auto}.  This term is indeed the dominant contribution to the total bias on large scales.  The error bars on the solid curves are computed from the scatter amongst the ten Gaussian kSZ realizations.}
\label{fig:results_auto_full_sims}
\end{figure}

The results of this procedure are shown in Figure~\ref{fig:results_auto_full_sims}.  As expected, the total bias for Planck remains negligible compared to the statistical errors, although there is a slight hint of a deficit at low-$L$.  For CMB-S3 and CMB-S4, Figure~\ref{fig:results_auto_full_sims} confirms that the term which we computed analytically (Equation~\ref{eq:auto}, i.e., Term B) is indeed the dominant term, particularly on large scales where the statistical errors are smallest.  For $\ell_{\rm max}=3000$ reconstructions, this term alone essentially suffices to describe the full bias, although there is a hint of an additional contribution around $L\approx 1500$.  For $\ell_{\rm max}=4000$ reconstructions, the contribution of the additional terms (Term C, due to the ``secondary contraction'', and Term E, due to the kSZ trispectrum) can clearly be seen.  These terms partially cancel the bias due to Term B on large scales, leading to a total bias that is slightly smaller in amplitude than Term B alone.  Term B dominates the total bias up to $L \approx 2500$; it appears that the other terms dominate the total bias on very small scales, although this is of less interest due to the larger statistical errors there.  Most importantly, the total bias is still significantly larger than the statistical errors for CMB-S3 and CMB-S4 temperature lensing reconstruction, as can be seen by comparing the solid curves in Figure~\ref{fig:results_auto_full_sims} to the expected error bars shown in Figure~\ref{fig:results_auto}.  Thus, mitigation strategies for the kSZ-induced bias will be needed for these experiments.


\section{Mitigation Strategies}
\label{sec:mitigation}

We have shown that the kSZ effect leads to significant biases in both the auto- and cross-power spectra of reconstructed CMB lensing maps. Here we discuss methods to reduce or eliminate the impact of these biases.  Unfortunately, all strategies described here come at the cost of decreased statistical significance in the lensing reconstruction.

\subsection{Polarization reconstruction}
To lowest order (in both galaxy optical depth and velocity), the kSZ effect produces only temperature anisotropies, not polarization anisotropies.  Thus, it only affects lensing reconstruction from CMB temperature maps; polarization-only reconstruction is free from the kSZ-induced biases discussed in this paper.  However, for map noise levels $\gtrsim 5 \mu$K-arcmin, temperature-based lensing reconstruction has larger statistical power than polarization reconstruction. Thus, polarization-only reconstruction would lead to a large degradation in $S/N$~\cite{2002ApJ...574..566H}, particularly for Stage-3 experiments.  For example, for the fiducial CMB-S3 configuration assumed here, the temperature estimator contributes about 75\% of the total $S/N$ on the CMB lensing auto-power spectrum measurement~\cite{LiuHill2016}.  The existence of the kSZ bias could thus motivate Stage-4 lensing survey designs that are optimized for depth (i.e., lower noise level) rather than large sky area, so that the polarization estimators dominate the lensing reconstruction $S/N$.  However, for CMB ``halo lensing'' measurements~\cite{Madhavacheril2015,Baxter2015,Melin2015} (i.e., the one-halo term of stacked CMB lensing measurements on a given halo sample), the temperature estimator is likely to always have higher $S/N$ than the polarization estimators (modulo foreground complexities), due to the much larger temperature gradient signal. The kSZ bias will thus require careful treatment for halo lensing measurements (see~\cite{Baxter2015,Raghunathan2017} for initial work in this direction).

\subsection{Reducing the reconstruction $\ell_{\rm max}$}
The kSZ-induced bias can also be decreased by restricting the lensing reconstruction to larger angular scales, i.e., lower $\ell_{\rm max}$.  This is because the relative contribution of the kSZ signal to the CMB power spectrum increases at higher $\ell$, and becomes the dominant source of anisotropy at $\ell \gtrsim 4000$ (assuming all non-blackbody signals have been removed). In Figures~\ref{fig:results_cross} and \ref{fig:results_auto} we show a comparison between $\ell_{\rm max} = 4000$ and $\ell_{\rm max} = 3000$, while in our tests we also consider $\ell_{\rm max} = 2000$.\footnote{Note that the CMB-S4 Science Book assumes $\ell_{\rm max} = 5000$ for temperature-based lensing reconstruction (see their Figure 46)~\cite{2016arXiv161002743A}.}  We find that for a CMB-S4 like experiment in cross-correlation with LSST galaxy lensing, the maximum bias at low $L$ goes from $\approx 15$\% for $\ell_{\rm max} = 4000$ to 5\% and 0.4\% when $\ell_{\rm max} = 3000$ and 2000, respectively. Similarly, the maximum bias to the auto-power spectrum (from Term B only) is reduced from $\approx 8$\% when $\ell_{\rm max} = 4000$ to 3\% and 0.3\% when $\ell_{\rm max} = 3000$ and 2000, respectively.  Therefore, in order for the kSZ-induced biases to be less than $1$\% (if no other mitigation strategy is applied), we would need to take $\ell_{\rm max} \lesssim 2000$.  Note that when reducing $\ell_{\rm max}$, a non-negligible kSZ bias seems to appear at high $L$ (however, this could be within the statistical errors on these small scales).

A reduction in $\ell_{\rm max}$ comes at a significant statistical cost, as summarized in Table \ref{tab:SNR_S4}: reducing $\ell_{\rm max}$ from 4000 to 3000 or 2000 yields a decrease in $S/N$ of a factor of $1.5 - 2$ or $3 - 5$, respectively (where the decrease depends on the observable considered).  In particular, the CMB-S4 lensing auto-power spectrum $S/N$ (from temperature reconstruction only) is reduced by a factor of 5 when reducing $\ell_{\rm max}$ from 4000 to 2000.

\begin{table}[h]
\begin{center}
  \begin{tabular}{| c | c | c | c |}
   \hline 
     \ $S/N$ for CMB-S4 \  & \ $\ell_{\rm max}$ = 4000 \  & \ 3000 \  & \ 2000 \  \\ \hline \hline
    $\langle \delta_g\ \kc \rangle$ & 497 & 281 & 127   \\ \hline
    $\langle \kg \ \kc \rangle$ & 251 & 157 & 80   \\ \hline
    $\langle \kc \ \kc \rangle$ & 252 & 140 & 50  \\ \hline
 \end{tabular}
 \caption{Expected $S/N$ for cross-correlations between LSST and CMB-S4 and for the CMB-S4 lensing auto-power spectrum. The power spectra are always assumed to be measured on the same multipole range $L = 30 - 3000$, while the lensing reconstruction is performed on temperature multipoles between $\ell_{\rm min} = 30$ and $\ell_{\rm max} = 4000, 3000$, and 2000 (no polarization information is used). For LSST lensing, the shape noise is assumed to be $\sigma_{\epsilon} = 0.26$ and the source number density $n = 26$ arcmin$^{-2}$. Here, we are using $f_{\rm sky} = 0.5$ for CMB-S4 and $f_{\rm sky} = 0.44$ for the overlap between LSST and CMB-S4.}
  \label{tab:SNR_S4}
\end{center}
\end{table}

\subsection{Other strategies}
As discussed in~\cite{2004NewA....9..687A, 2014ApJ...786...13V}, masking the most massive galaxy clusters and brightest point sources can reduce lensing reconstruction biases due to astrophysical signals, including the kSZ and tSZ effects, as well as dust or radio emission.  Indeed, this strategy has been used to mitigate biases from the tSZ effect and point source emission in recent lensing analyses from Planck, ACT, and SPT~\cite{Planck2016lensing,Sherwin2016,Story2015}.  To reduce the kSZ bias discussed in this paper, galaxy groups and clusters must be masked.  In particular, the kSZ signal is proportional to the cluster mass, and thus by masking the most massive clusters (and then progressively decreasing the masking threshold), the kSZ-induced bias can be progressively decreased.  However, the lensing signal of these objects is also proportional to their mass.  Thus, this strategy is guaranteed to bias the $\hat{\kappa}_{\rm CMB}$ reconstruction itself, at some level, because the masked regions are preferentially the highest $\kc$ regions in the sky.  For current analyses, the effect of this ``high-mass masking'' is negligible on the CMB lensing power spectrum (compared to the statistical errors), but it may already be an issue for cross-correlations with galaxy lensing maps~\cite{LiuHill2015} and is clearly an issue for tSZ cross-correlations~\cite{Hill-Spergel2014}.   For our purposes here, it is unclear that a significant reduction of the kSZ-induced bias can be achieved by masking without simultaneously biasing the lensing reconstruction at an unacceptable level (particularly for low-redshift cross-correlations). Further numerical work would be required to explore this point.

Given the analytical templates produced in this paper, the amplitude of the kSZ contamination can be estimated together with the amplitude of lensing, thus greatly reducing the leakage of one signal into the other. This procedure, known as ``bias hardening'', was explored in \cite{2014JCAP...03..024O, 2013MNRAS.431..609N}.  However, this method would be complicated by the fact that the kSZ contribution depends sensitively on baryonic effects, as shown in Figures~\ref{fig:results_cross} and~\ref{fig:results_auto}.  Thus, the templates would come with additional theoretical uncertainty (unlike, e.g., templates appropriate for the trispectrum of Poisson-distributed point sources).  Furthermore, the bias-hardening would lead to some loss of $S/N$.  We leave such calculations for future work.

A final idea for mitigating the kSZ-induced bias relies on the approximate symmetries of the problem: the standard lensing quadratic estimator optimally combines estimates of the local dilation and shear of the background primary CMB~\cite{2012PhRvD..85d3016B, 2017arXiv170902227P, 1999PhRvD..59l3507Z}. Since the kSZ field at low redshift is mostly sourced by galaxies and clusters, we expect the kSZ signal to predominantly contaminate the ``dilation'' part of the lensing estimator.  Thus, we speculate that a ``shear''-only reconstruction would be less affected by kSZ contamination.\footnote{However, the large-scale tidal component of the density field will also contribute to the shear; thus, a shear-only estimator could still receive a small kSZ contribution.} Similarly, one could construct an estimator sensitive only to kSZ by taking the appropriate difference of dilation-only and shear-only lensing estimators, since lensing contributes in the same way to both (up to a factor), while the kSZ signal contributes differently.  In addition, real-space estimators have been proposed that are sensitive to only kSZ, and not lensing, due to the conservation of surface brightness by lensing~\cite{Riquelme2007}.  A full exploration of such avenues is left to future work.

Lastly, we note that kSZ contamination could also lead to a failure of the usual curl null test in CMB lensing reconstruction.  Since the kSZ field is not the gradient of a scalar field (unlike the CMB lensing deflection), it will generically yield a non-zero curl reconstruction.  This test can thus be used as a diagnostic for kSZ contamination, although other systematics and foregrounds can also contribute to the curl, which may render the test non-informative as to the origin of the failure.

\section{Conclusions}
\label{sec:conclusions}
CMB lensing measurements from ongoing and upcoming experiments will be one of the most powerful cosmological probes available in the near term. The CMB lensing power spectrum measures the amplitude of late-time matter fluctuations over a broad range of redshifts and is sensitive to a variety of novel physics, including massive neutrinos, dark energy, and modified gravity.  At the same time, cross-correlations of the CMB lensing field with low-redshift tracers (such as galaxy number density or galaxy lensing convergence maps) in several redshift bins can probe the time evolution of the matter fluctuations, breaking degeneracies between different models and allowing further improvements in cosmological constraints, especially for non-standard models.

However, lensing reconstruction is afflicted by biases related to non-Gaussian-distributed astrophysical sources (which are themselves generally correlated with the lensing field).  Here, we have focused on the kSZ effect, which is the largest contaminant that cannot be removed via multifrequency component separation techniques, since the kSZ effect preserves the blackbody spectrum of the CMB.  We have shown that for an aggressive reconstruction with $\ell_{\rm max} = 4000$, the biases to cross-correlations with LSST lensing maps can be as large as $\approx$ 2\%, 12\%, and 15\% for CMB experiments similar to Planck, CMB-S3, and CMB-S4, respectively.  The biases to CMB lensing auto-power spectrum measurements can be as large as $\approx$ 1\%, 6\%, and 8\% for Planck, CMB-S3, and CMB-S4, respectively, when using $\ell_{\rm max} = 4000$, and about half of that for $\ell_{\rm max} = 3000$.  Moreover, the kSZ-induced bias has non-negligible sensitivity to the assumptions made about the baryon distribution, making it difficult to predict \emph{ab initio}, as seen in the differences between the analytic and simulation-derived results in our work.  For Planck, the bias is smaller than the statistical error bars on the lensing power spectrum.  However, the kSZ-induced bias is considerably larger than the statistical precision of Stage 3 and 4 CMB experiments, and is larger than the few-percent change induced on the lensing auto-power spectrum by massive neutrinos.  Thus, it will require careful consideration in future analyses.  We have verified the amplitude of these effects by comparing directly to measurements from cosmological simulations, including the first full simulation-based calculation of a secondary-induced CMB lensing bias (i.e., including all terms).  Nevertheless, precise predictions of the kSZ-induced biases will require simulations with more sophisticated baryonic feedback implementations than those considered here.

Mitigation strategies to reduce this bias include the use of polarization-only reconstruction or the reduction of the maximum temperature multipole $\ell_{\rm max}$ used in the lensing reconstruction. In order to ensure that the bias is always less than 1\% on large scales, we find that we would need to take $\ell_{\rm max} \lesssim 2000$, which would lead to a reduction in statistical $S/N$ on various observables by a factor of $3-5$ for CMB-S4. Other strategies such as masking, building bias-hardened estimators, or using shear-only reconstruction will be the subject of future work.

Finally, we note that in a realistic experiment, imperfect foreground removal can introduce additional biases, for example from residual tSZ or CIB~\cite{2014ApJ...786...13V,2014JCAP...03..024O}. The exact size of these residuals depends on the experimental configuration, the multifrequency component separation method, and the true complexity of the small-scale microwave sky (e.g., possible decoherence of the CIB across frequencies).  The residuals may lead to biases that are larger than or comparable to the kSZ-induced bias discussed in this work --- indeed, if no multifrequency cleaning or masking were performed (e.g., at 150 GHz), they would be larger than the kSZ bias.  Nevertheless, in principle the other biases can be removed at high precision with sensitive measurements at multiple frequencies, whereas the kSZ bias cannot be.

\acknowledgments
We are grateful to Shirley Ho, Mathew Madhavacheril, Emmanuel Schaan, Uro\^{s} Seljak, Blake Sherwin, Kendrick Smith, David Spergel, and Alexander van Engelen for useful conversations and comments.  We also thank the anonymous referee for comments that substantially improved the manuscript.  SF thanks the Miller Institute for Basic Research in Science at the University of California, Berkeley for support.  This work was partially supported by a Junior Fellow award from the Simons Foundation to JCH.  Some of the results in this paper have been derived using the HEALPix package~\cite{Gorskietal2005}.

\begin{appendix}
\section{Derivation of kSZ bias to CMB lensing-tracer cross-correlation}
\label{app:cross}
Here, we compute the kSZ$^2$ contamination to $\langle \delta_g \ \hat{\kappa}_{\rm CMB} \rangle$ or $\langle \kappa_{\rm gal} \ \hat{\kappa}_{\rm CMB} \rangle$ and compare to the fiducial signals produced by lensing only.  We consider temperature reconstruction only, since polarization is much less affected by the kSZ signal.

The quadratic estimator for CMB lensing in terms of the lensed CMB temperature fluctuations $\tTheta$ can be written as
\be
\hat{\kappa}_{\rm CMB}(\bl) = \frac{L^2 N(\bl)}{2} \int_{\L} \tTheta(\L) \tTheta(\bl - \L) f(\L, \bl) = \int_{\L_1} \int_{\L_2} \tTheta(\L_1) \tTheta(\L_2) \ (2 \pi)^2 \delta_D(\bl - \L_1 - \L_2) \Gamma(\L_1, \bl) \,,
\ee
where we have defined 
\be
\Gamma(\L, \bl) = \frac12 L^2 N(\bl) f(\L, \bl) 
\ee
and 
\be
f(\L, \bl) = \frac{ (\bl - \L) \cdot \bl C^{TT}_{|\bl - \L|} + \L \cdot \bl C^{TT}_{\ell} }{ 2C^{\rm tot}_{\ell}  C^{\rm tot}_{|\bl - \L|} }\ \ , \ \ \ \ \ \ \  N(\bl)^{-1} = \int_\L \frac{ \left[(\bl - \L) \cdot \bl C^{TT}_{|\bl - \L|} + \L \cdot \bl C^{TT}_{\ell} \right]^2 }{ 2C^{\rm tot}_{\ell}  C^{\rm tot}_{|\bl - \L|} } \,.
\ee

As we argued in Section \ref{sec:cross}, to calculate the kSZ bias to CMB lensing, one can simply replace $\tTheta \rightarrow \Theta^{\rm kSZ}$ in the $\hat{\kappa}_{\rm CMB}$ estimator. For the kSZ field we can write:
\be
\Theta^{\rm kSZ}(x, y) = - \int d\eta \ g(\eta) \ p_z(x, y, \eta) \,,
\ee
where $x$ and $y$ are (angular) displacements across the line-of-sight, $\eta$ is the comoving distance in the $z$ direction, $g(\eta) = \dot{\tau} e^{-\tau}$ is the visibility function, and the line-of-sight momentum $p_z \approx \delta_e v_z$ on small scales. 
Taking the Fourier transform, we find that the projected kSZ temperature fluctuation is:
\be
\Theta^{\rm kSZ}(\L)  = - \int \frac{d \eta}{\eta^2} g(\eta) \int  \frac{d k_z}{2\pi} \ \tilde{p}_z(\k_{\perp} = \L / \eta, k_z) \ e^{i k_z \eta} \,,
\ee
where $\tilde{p}_z(\k)$ is the Fourier transform of $p_z(x, y, \eta)$.  Similarly, for the projected galaxy fluctuation (or galaxy lensing convergence) we have:
\be
\delta_g(x, y) =  \int d\eta \ W^g(\eta) \ \delta(x, y, \eta)
\ee
so that in Fourier space
\be
\delta_g(\bl)  = \int \frac{d \eta}{\eta^2} W^g(\eta) \int  \frac{d k_z}{2\pi} \ \tilde{\delta}_g (\k_{\perp} = \bl / \eta, k_z) \ e^{i k_z \eta} \,.
\ee

Now we can compute the kSZ$^2$ bias to the CMB lensing-tracer cross-correlation, $\langle \delta_g(\bl_1) \hat{\kappa}_{\rm CMB}^{\rm kSZ}(\bl_2) \rangle$:
\be
\langle \delta_g(\bl_1) \hat{\kappa}_{\rm CMB}^{\rm kSZ} (\bl_2) \rangle = \int_{\L_3, \L_4} \Gamma(\L_3, \bl_2) \langle \delta_g(\bl_1)  \Theta^{\rm kSZ}(\L_3) \Theta^{\rm kSZ}(\L_4) \rangle \ (2\pi)^2 \delta_D(\bl_2 - \L_3 - \L_4)
\label{eq:cross_explicit}
\ee

Renaming indices, the expectation value in the integrand is given by:
\ba
\langle \delta_g(\bl_1) \Theta^{\rm kSZ}(\L_2) \Theta^{\rm kSZ}(\L_3) \rangle &=& \int\displaylimits_{\eta_1,\eta_2,\eta_3} \int\displaylimits_{k_{1z}, k_{2z}, k_{3z}}
 e^{i(k_{1z}\eta_1 +k_{2z}\eta_2+k_{3z}\eta_3)} \frac{W^g(\eta_1) g(\eta_2) g(\eta_3)}{\eta_1^2 \eta_2^2 \eta_3^2} \langle \delta_g(\k_1) p_z(\k_2) p_z(\k_3) \rangle \nn \\
& &\hskip -5cm = \int\displaylimits_{\eta_1,\eta_2,\eta_3} \int\displaylimits_{k_{1z}, k_{2z}, k_{3z}} e^{i(k_{1z}\eta_1 +k_{2z}\eta_2+k_{3z}\eta_3)} \frac{W^g(\eta_1) g(\eta_2) g(\eta_3)}{\eta_1^2 \eta_2^2 \eta_3^2} (2 \pi)^3\delta_D(\k_{1\perp} +\k_{2\perp} + \k_{3\perp}) \delta_D(k_{1z} + k_{2z} + k_{3z}) B_{\delta p_z p_z}(\k_1, \k_2, \k_3) \nn \\
& &\hskip -5cm = \int\displaylimits_{\eta_1,\eta_2,\eta_3} \int\displaylimits_{k_{1z}, k_{3z}} e^{i k_{1z}(\eta_1 -\eta_2)}  e^{i k_{3z}(\eta_3 -\eta_2)}   \frac{W^g(\eta_1) g(\eta_2) g(\eta_3)}{\eta_1^2 \eta_2^2 \eta_3^2} (2 \pi)^2\delta_D(\k_{1\perp} +\k_{2\perp} + \k_{3\perp})  B_{\delta p_z p_z}(\k_1, \k_2, \k_3)
\ea
So far the result is exact. We can now use the Limber approximation, treating the integrand as slowly varying in $\eta$ and doing the $k_z$ integrals:
\ba
\hskip -0.5cm \langle \delta_g(\bl_1) \Theta^{\rm kSZ}(\L_2) \Theta^{\rm kSZ}(\L_3) \rangle  \hskip -0.1cm &\approx& \hskip -0.6cm \int\displaylimits_{\eta_1,\eta_2,\eta_3}  \hskip -0.4cm \delta_D(\eta_1 - \eta_2) \delta_D(\eta_3 - \eta_2) \frac{W^g(\eta_1) g(\eta_2) g(\eta_3)}{\eta_1^2 \eta_2^2 \eta_3^2} (2 \pi)^2\delta_D(\k_{1\perp} +\k_{2\perp} + \k_{3\perp})  B_{\delta p_z p_z}(\k_1, \k_2, \k_3) \nonumber \\
& = & \int_{\eta_1} \frac{W^g(\eta_1) g^2(\eta_1)}{\eta_1^4} (2 \pi)^2\delta_D(\bl_1 + \L_2 + \L_3)  B_{\delta p_z p_z} \left( \k_1 = \frac{\bl_1}{\eta_1}, \k_2 = \frac{\L_2}{\eta_1}, \k_3 = \frac{\L_3}{\eta_1} ; \eta_1 \right)
\ea
Here, the hybrid bispectrum arises from momenta lying on surfaces of constant redshift at distance $\eta_1$. \\

Switching back indices $2 \rightarrow 3$ and $3 \rightarrow 4$, and plugging this into Equation~\ref{eq:cross_explicit}, we find:
\ba
\langle \delta_g(\bl_1) \hat{\kappa}_{\rm CMB}^{\rm kSZ}(\bl_2) \rangle &=& \int_{\L_3, \L_4} \Gamma(\L_3, \bl_2) \langle \delta_g(\bl_1)  \Theta^{\rm kSZ}(\L_3) \Theta^{\rm kSZ}(\L_4) \rangle \ (2\pi)^2 \delta_D(\bl_2 - \L_3 - \L_4) \nn \\
 &  & \hskip -5cm = \int_{\eta_1} \frac{W^g(\eta_1) g^2(\eta_1)}{\eta_1^4} \int_{\L_3, \L_4} \Gamma(\L_3, \bl_2) (2\pi)^2 \delta_D(\bl_1 + \L_3 + \L_4) \ (2\pi)^2 \delta_D(\bl_2 - \L_3 - \L_4)  B_{\delta p_z p_z}(\k_1, \k_3, \k_4) \nn \\
  &  & \hskip -5cm =(2\pi)^2 \delta_D(\bl_1 + \bl_2) \int_{\eta_1} \frac{W^g(\eta_1) g^2(\eta_1)}{\eta_1^4} \int_{\L_3} \Gamma(\L_3, \bl_2) B_{\delta p_z p_z}(\k_1, \k_3, \k_2 - \k_3)
\ea
This is the final result.  If we want to put it in a more familiar form, we can change variables to $\q = \L_3 / \eta$ (note that all momenta are perpendicular to the line-of-sight) and use the fact that $d^2\L_3 = \eta^2 d^2\q$ to find the expression for the kSZ-induced bias given in Equation~\ref{eq:cross} of the main text:
\ba
\left( \Delta C_L^{\kappa_{\rm CMB} \times g} \right)_{\rm kSZ} & = & \int_{\eta} \frac{W^g(\eta) g^2(\eta)}{\eta^2} \int_\q \Gamma(\L_3 = \q \eta, \bl_2 = - \bl) B_{\delta p_z p_z}(\k = \bl/\eta, \q, -\k - \q; \eta) \nonumber \\
& = & \int_{\eta} \frac{W^g(\eta) g^2(\eta)}{\eta^2} \int_\q \Gamma(\bl + \q \eta, \bl) B_{\delta p_z p_z}(\k = \bl/\eta, \q, -\k - \q; \eta) \,,
\label{eq:final_cross}
\ea
where we can approximate (see \cite{Doreetal2004, DeDeoetal2005})
\be
B_{\delta p_z p_z} \approx \frac13 v_{\rm rms}^2 B_m \,.
\ee
The same result holds if we consider galaxy lensing instead of galaxy overdensity, with the replacement $W^g \rightarrow W^{\kappa_{\rm gal}}$.

\section{kSZ bias to CMB lensing auto-power spectrum}
\label{app:auto}

Here, we discuss the lowest-order bias to the $\hat{\kappa}_{\rm CMB}$ power spectrum due to the kSZ effect.  In analogy to Equation~\ref{eq:cross_explicit}, we can write:
\be
\langle \hat{\kappa}_{\rm CMB}(\bl_1) \hat{\kappa}_{\rm CMB}(\bl_2) \rangle = \int_{\L_3, \L_4,  \L_5,  \L_6} \Gamma(\L_4, \L_3) \Gamma(\L_6, \L_5) \ \langle \tTheta(\L_3) \tTheta(\L_4) \tTheta(\L_5) \tTheta(\L_6) \rangle \ (2\pi)^4 \delta_D(\bl_1 - \L_3 - \L_4) \delta_D(\bl_2 - \L_5 - \L_6)
\label{eq:auto_explicit}
\ee

We decompose the observed fluctuations $\tTheta = \tTheta_p + \Theta^{\rm kSZ}$ and organize the ensemble average in Equation \ref{eq:auto_explicit}, $\langle \tTheta(\L_3) \tTheta(\L_4) \tTheta(\L_5) \tTheta(\L_6) \rangle$, as the sum of terms of the form summarized in Table \ref{tab:terms}.

\begin{table}[h]
\begin{center}
  \begin{tabular}{| c | c | c | c |}
    \hline 
     label & multiplicity & form & notes  \\ \hline \hline
    A & 4 & $\langle \Theta^{\rm kSZ} \tTheta_p \tTheta_p \tTheta_p \rangle$  & vanishes due to symmetry \\ \hline
    B & 2 & $\langle (\Theta^{\rm kSZ} \Theta^{\rm kSZ}) (\tTheta_p \tTheta_p) \rangle$  & same as $\langle \kappa_{\rm CMB}^{\rm kSZ} \kappa_{\rm CMB}^{\rm true}  \rangle$  \\ \hline
    C & 4 & $\langle (\Theta^{\rm kSZ} \tTheta_p) (\Theta^{\rm kSZ} \tTheta_p) \rangle$  & ``secondary contraction'', see discussion below \\ \hline
    D & 4 & $\langle \tTheta_p \Theta^{\rm kSZ} \Theta^{\rm kSZ} \Theta^{\rm kSZ} \rangle$  & vanishes due to symmetry \\ \hline
    E & 1 & $\langle \Theta^{\rm kSZ} \Theta^{\rm kSZ} \Theta^{\rm kSZ} \Theta^{\rm kSZ} \rangle$  & kSZ trispectrum  \\ \hline
    F & 1 & $\langle \tTheta_p \tTheta_p \tTheta_p \tTheta_p \rangle$  & same as $\langle \kappa_{\rm CMB}^{\rm true} \kappa_{\rm CMB}^{\rm true}  \rangle$ after noise bias subtraction \\ \hline

  \end{tabular}
  \caption{Different terms in the expansion of the expectation value in Equation \ref{eq:auto_explicit}. The multiplicity denotes the combinatorial factor in the expansion, and the parentheses ( ) make explicit which variables are convolved together. Note that all of the quantities $\Theta_p$, $\kappa_{\rm CMB}$, and $\Theta^{\rm kSZ}$ are $\ll 1$ and therefore we shall only keep the ones with the fewest number of powers of the small quantities.}
  \label{tab:terms}
\end{center}
\end{table}
The first thing to note is that any Gaussian component of the fields $\tTheta_p$ and $\Theta^{\rm kSZ}$ will be subtracted when removing the noise bias $N^{(0)}$ in the process of estimating the $\kc$ power spectrum from the measured (biased) $\hat{\kappa}_{\rm CMB}$ power spectrum.  Therefore, we will neglect all terms that arise from Wick's theorem (i.e., any expansion of higher (even-order) point functions as a product of two-point functions). 

Also, terms containing an odd power of $\Theta^{\rm kSZ}$ vanish on average because the kSZ effect is equally likely to be positive or negative (a consequence of the symmetry $v_r \rightarrow -v_r$).

Term B corresponds to estimating $\hat{\kappa}_{\rm CMB}$ on the kSZ field alone and then cross-correlating with the true $\kc$. This is equivalent to the calculation in the previous section, where now $\kappa_{\rm CMB}$ is the low-redshift matter tracer rather than galaxies. Therefore, this term can be computed by replacing $W^g \rightarrow W^{\kc}$ in Equation \ref{eq:final_cross} above (and noting the multiplicity of 2 in this case).

Evaluating Term C is equivalent to applying the lensing quadratic estimator on two different maps, where one leg is taken from $\tTheta_p$ and the other from $\Theta^{\rm kSZ}$, and then evaluating the resulting power spectrum. This represents a ``secondary contraction'' (in the terminology of~\cite{2014JCAP...03..024O}) of the same fields that give rise to Term B, and is analogous to the $N^{(1)}$ bias for the CMB lensing field itself~\cite{Kesden2003}. Such ``secondary contraction'' terms have been investigated for the biases arising from Poisson radio point sources, tSZ clusters, and Poisson and clustered infrared sources~\cite{2014JCAP...03..024O}, and it has been shown that these terms can be of the same order of magnitude as the ``primary contraction'' (Term B), particularly at high $L$ for high-resolution, low-noise experiments. It is non-trivial to evaluate Term C analytically, since the full expression results in non-separable integrals over the bispectrum.


Term E is only sourced by the (connected part of the) kSZ trispectrum. The contribution from late-time matter has been shown to be roughly 10 times smaller than Term B for lensing reconstruction with noise levels matching those of the original ACT survey and $\ell_{\rm max} \approx 2000$ (after $N^{(0)}$ subtraction)~\cite{2011PhRvL.107b1301D}.  For the lower noise levels of CMB-S3 and CMB-S4 and/or higher $\ell_{\rm max}$ values, Term E and Term B could be more comparable in amplitude, but quantifying this analytically is beyond the scope of the present paper.  The sum of Term B and Term E was also considered in~\cite{2012ApJ...756..142V}, who found a combined sub-percent bias on the CMB lensing auto-power spectrum for noise levels matching those of the SPT-SZ survey and $\ell_{\rm max} = 3000$.

We present numerical results for the total bias (i.e., the sum of Terms B+C+E) in Sec.~\ref{sec:fullauto}.  Term B is indeed the dominant term, particularly on large scales.  For $\ell_{\rm max} = 4000$, cancellations amongst the various terms can clearly be seen.

All of our results have neglected the kSZ signal from reionization. We note that Ref.~\cite{2016arXiv160701769S} has argued that the kSZ signal due to fluctuations in the ionization fraction during reionization can cause a detectable squeezed limit trispectrum.  We leave the study of the impact of such contributions on CMB lensing to future work.

\end{appendix}

\end{document}